\documentclass[preprint2]{aastex}
\usepackage{graphicx}
\usepackage{natbib}
\usepackage{lscape}

\shorttitle{T\,Tauri Jet Rotation at NUV and Optical Wavelengths}
\shortauthors{Coffey et al. 2007}

\begin{document}

\title{Further Indications of Jet Rotation in New Ultraviolet and Optical {\em HST}/STIS Spectra
\footnote{Based on observations made with the 
NASA/ESA $\it{Hubble}$ $\it{Space}$ $\it{Telescope}$, 
obtained at the Space Telescope Science Institute, which is operated by the Association of 
Universities for Research in 
Astronomy, Inc., under NASA contract NAS5-26555.}}

\author{
Deirdre Coffey\altaffilmark{1}, Francesca Bacciotti\altaffilmark{1}, Thomas P. Ray\altaffilmark{2}, 
Jochen Eisl\"{o}ffel\altaffilmark{3}, Jens Woitas\altaffilmark{3}}

\altaffiltext{1}{I.N.A.F. - Osservatorio Astrofisico di Arcetri, Largo E. Fermi 5, 50125 
Firenze, Italy \email{dac, fran@arcetri.astro.it}}
\altaffiltext{2}{Dublin Institute for Advanced Studies, 5 Merrion Square, Dublin 2, 
Ireland \email{tr@cp.dias.ie}}
\altaffiltext{3}{Th\"{u}ringer Landessternwarte Tautenburg, Sternwarte 5, 07778 
Tautenburg, Germany \email{jochen@tls-tautenburg.de}}

\begin{abstract}
We present survey results which suggest rotation signatures at the base of T\,Tauri jets. Observations were conducted with the Hubble Space Telescope 
Imaging Spectrograph at optical and near ultraviolet wavelengths (NUV). 
Results are presented for the approaching jet from DG\,Tau, CW\,Tau, HH\,30 and the bipolar jet from TH\,28. Systematic asymmetries in Doppler shift were detected across the jet, within 100\,AU from the star. At optical wavelengths, radial velocity differences were typically 10\,to\,25~($\pm$5)\,km\,s$^{-1}$, while differences in the NUV range were consistently lower at typically 10~($\pm$5)\,km\,s$^{-1}$. 
Results are interpreted as possible rotation signatures. Importantly, there is agreement between the optical and NUV results for DG\,Tau. Under the assumption of steady magnetocentrifugal acceleration, the survey results lead to estimates for the distance of the jet footpoint from the star, and give values consistent with earlier studies. In the case of DG\,Tau, for example, we see that the higher velocity component appears to be launched from a distance of 0.2~to~0.5\,AU from the star along the disk plane, while the lower velocity component appears to trace a wider part of the jet launched from as far as 1.9\,AU. The results for the other targets are similar. Therefore, if indeed the detected Doppler gradients trace rotation within the jet then, under the assumption of steady MHD ejection, the derived footpoint radii support the existence of magnetized disk winds. However, since we do not resolved the innermost layers of the flow, we cannot exclude the possibility that there also exists an X-wind or stellar wind component.
\end{abstract}



\section{Introduction}

Collimated jets associated with young stars (\protect\protect\citealp{Eisloffel00}; \protect\protect\citealp{Bally07}) are believed to play a key role in the star formation process, 
as they may be able to extract a substantial amount of the excess angular momentum from the star/disk system
(e.g. \protect\protect\citealp{Ray03}; \protect\protect\citealp{Ray07}). Several models propose apparently feasible mechanisms for launching these jets 
(e.g. \protect\citealp{Ferreira97}; \protect\citealp{Shu00}; \protect\citealp{Pudritz07}; \protect\citealp{Shang07}) but the observational evidence favouring one model 
over another has remained elusive. Difficulties arise because of the small spatial scales involved (since jet launching 
occurs within a few AU of the star), and from the fact that many such protostars are heavily embedded. However, in recent 
years observations of a small number of jets emitting in H$_2$ lines, and of jets from T\,Tauri stars which can be 
traced back to the source, have provided new and interesting insights.

In particular, a new parameter for constraining theoretical models has emerged from recent high angular resolution 
observations: indications of {\em rotation} within these jets about their symmetry axis. This interpretation followed from the detection 
of systematic radial velocity differences across the jet in high angular resolution spectra of the jet borders taken with 
the slit placed parallel to the jet axis (\protect\citealp{Davis00}; \protect\citealp{Bacciotti02}; \protect\citealp{Woitas05}). More recently, we have 
undertaken a survey with the Hubble Space Telescope Imaging Spectrograph ({\em HST}/STIS) of eight jets, each from one of six pre-main sequence stars, in order to determine if signatures of rotation are common among these objects. 
For this study, performed at optical wavelengths, we adopted an instrument orientation in which the slit was perpendicular to the outflow direction. Relative to the parallel-slit orientation, this considerably reduced the exposure time needed to extract a possible rotation signature. The results for the first phase of observations in this survey, which included the bipolar jets from RW\,Aur and TH\,28, showed gradients in Doppler shifts across the jet of 10~to~25\,($\pm$\,5)\,km\,s$^{-1}$ \protect\cite{Coffey04}. A number of theoretical analyses then followed demonstrating the implications of interpreting measured Doppler gradients as jet rotation. That is, when combined with other jet parameters, measured radial velocities differences allow the extraction of important information such as the jet launching radius, under the assumption of steady magnetohydrodynamic (MHD) ejection (\protect\citealp{Anderson03}; \protect\citealp{Pesenti04}; \protect\citealp{Woitas05}; \protect\citealp{Ferreira06}). 

While the results of the first phase of our survey were very encouraging, the limiting resolution (spatially and spectrally) 
of {\em HST}/STIS in the optical wavelength range restricts us to the peripheral regions of the jet. However, emission from 
protostellar outflows is dominated by lines in the ultraviolet. In particular, models \protect\cite{Hartigan87} and 
observations \protect\cite{Hartigan99} of shocks both show that low excitation \ion{Mg}{2} doublet emission at 2796\,\AA\,\,and 
2803\,\AA\,\,is stronger in protosteller outflows than `traditionally' observed optical lines such as the 
[\ion{O}{1}]$\lambda\lambda$6300,6363 doublet. Therefore, {\em HST}/STIS observations were scheduled to examine the same targets at 
{\em near ultraviolet} (NUV) wavelengths. Three jet targets, of the five allocated, were observed before the failure of the 
{\em HST}/STIS power supply on August 3rd 2004.

We report here on the results of the second phase of observations for our optical survey, and the observations of the 
NUV survey. We present a radial velocity analysis of the optical dataset, which constitutes three jet targets i.e. the 
approaching jets from T\,Tauri stars DG\,Tau, CW\,Tau and HH\,30. We also present the NUV dataset, which also constitutes three jet 
targets but from two T\,Tauri stars, i.e. the approaching jet from DG\,Tau and the bipolar jet from TH\,28. Our 
observations are described in Section\,\protect\ref{observations}, while the method of analysis and results are detailed in 
Section\,\protect\ref{results}. In Section\,\protect\ref{discussion}, we discuss the measured parameters in the context of the 
magneto-centrifugal acceleration process, and we summarise our conclusions in Section\,\protect\ref{conclusions}. 

\section{Observations}
\label{observations}

\begin{figure*}
\begin{center}
\includegraphics[scale=0.35]{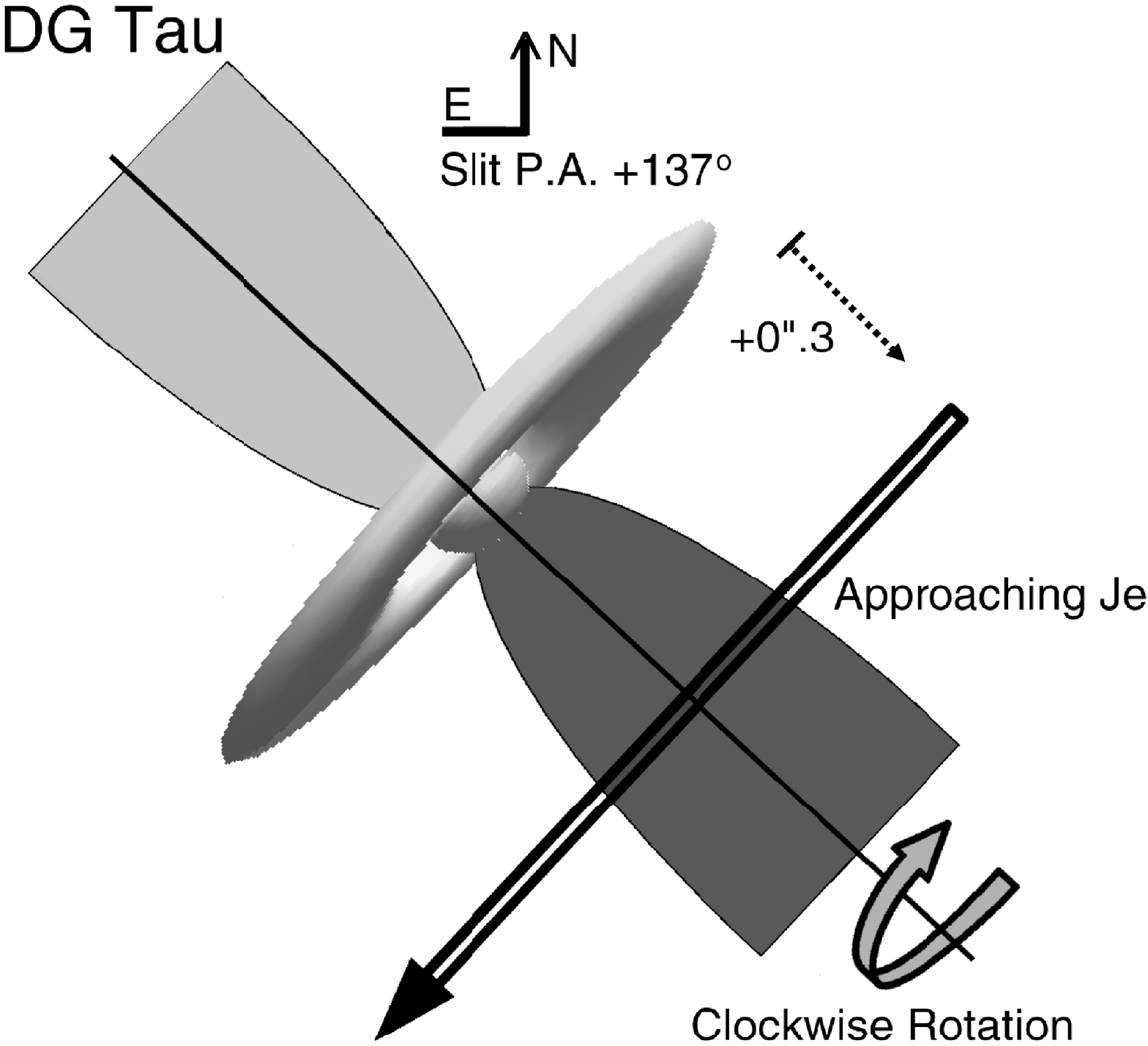}\includegraphics[scale=0.38]{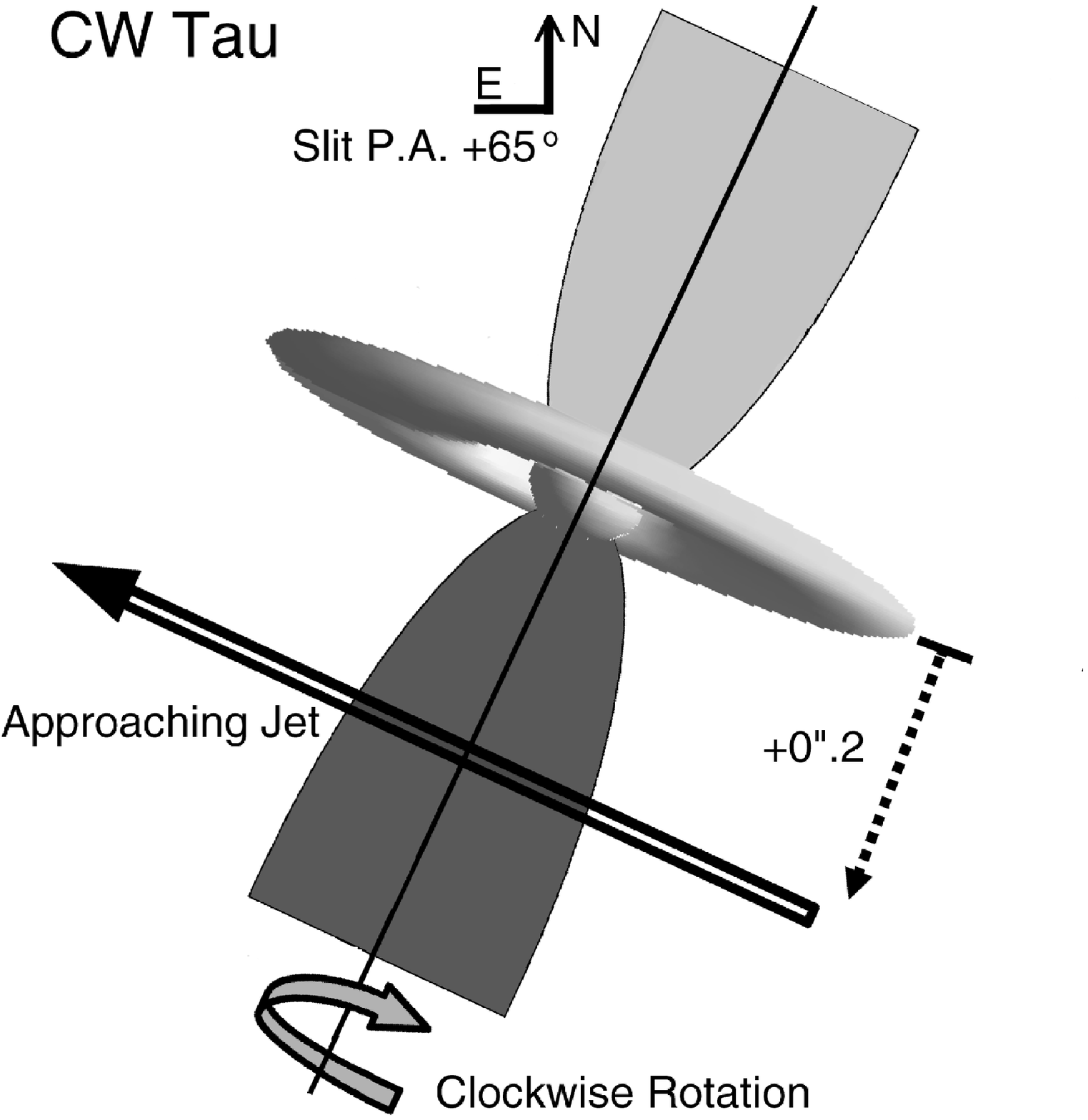} \\
\hspace{0.25 cm}\includegraphics[scale=0.38]{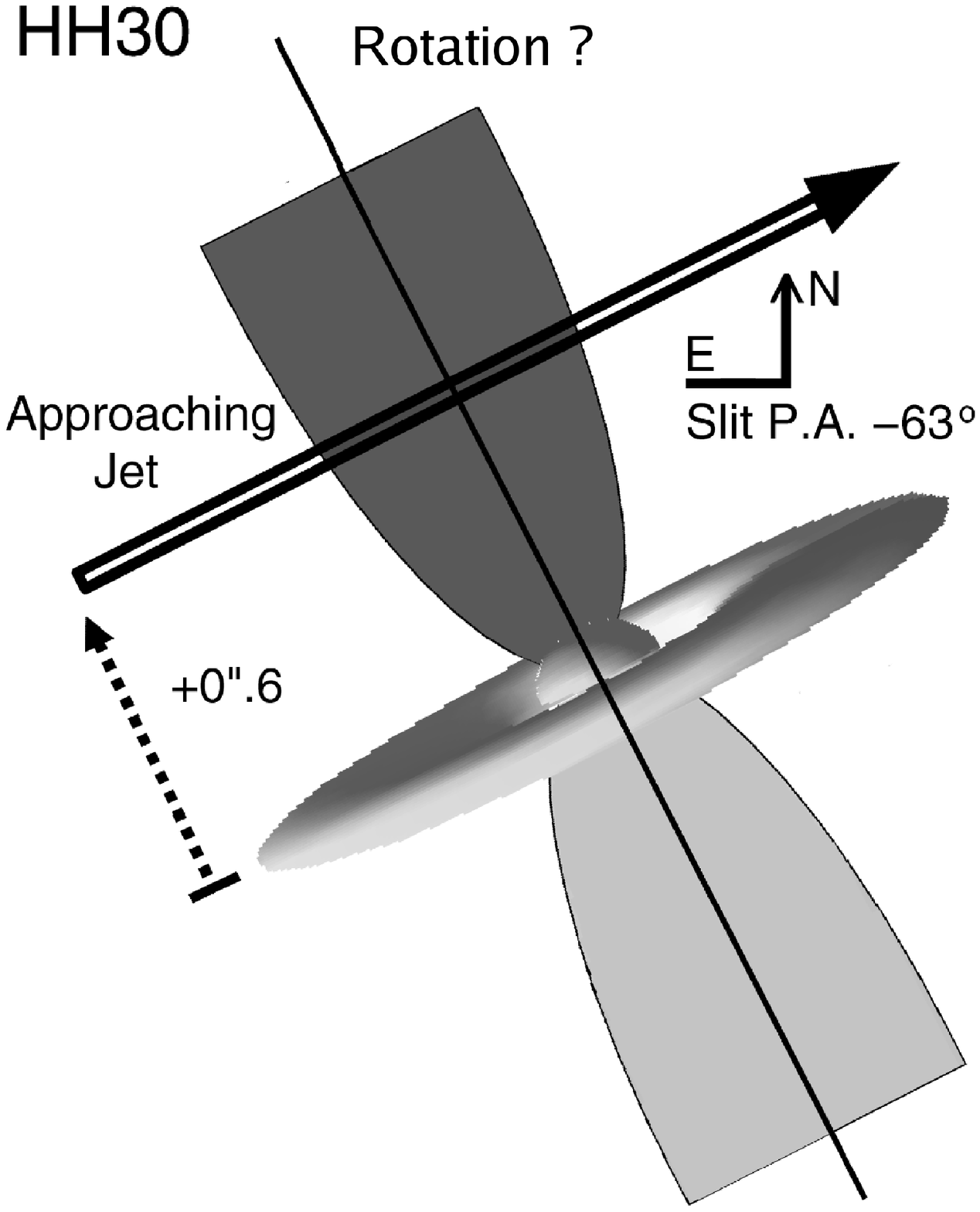}\includegraphics[scale=0.35]{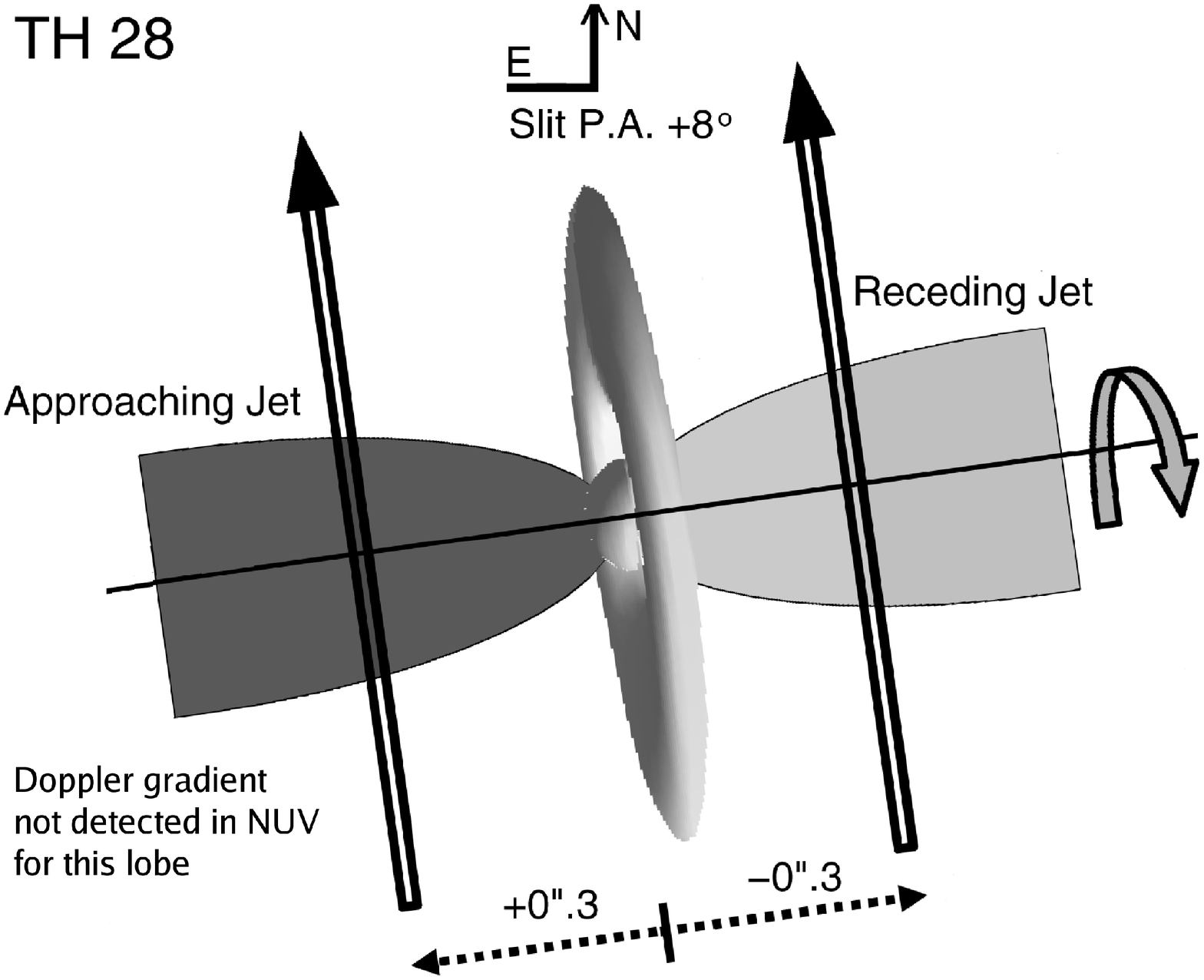}
\figcaption{Orientation of the jet and slit for each target. The arrow on the slit indicates the positive direction of 
the y-axis on the position-velocity contour plot, Figure\,\protect\ref{pv_optical_jets}. The jet orientation can be compared with compass orientation given in the upper corners of each radial velocity profile, Figures\,\protect\ref{velocityprofiles_opt} and \protect\ref{velocityprofiles_optuv}. The arrow around the jet axis indicates the sense of rotation as implied by the results in this paper. The requested slit position angle was 90$^{\circ}$ with respect to the value of PA$_{jet}$, Table\,\protect\ref{targets}. The actual slit position angles for DG\,Tau and HH\,30 differ from the requested values by +3$^{\circ}$ and -6$^{\circ}$ respectively. This was due to problems during observations in finding the right combination of guide stars for the requested angle. The sense of rotation is denoted according to the viewpoint of the observer looking down the approaching jet towards the star. 
\label{orientation}}
\end{center}
\end{figure*}

Spectroscopic observations, in both the optical and NUV wavelength regions, were made at the base of jets from several 
T\,Tauri stars (listed in Table\,\protect\ref{targets}). Jet position angles were determined from archival {\em HST} images. The observing procedure used involved centering the {\it HST}/STIS slit on the 
T\,Tauri star, rotating the slit to a position angle perpendicular to the jet axis, and then offsetting the slit to a position along the jet which is a fraction of an arcsecond from the source, as illustrated in Figure\,\protect\ref{orientation}.

\begin{table*}
\begin{center}
\scriptsize{\begin{tabular}{llcccccc}
\tableline\tableline
Target		&Location	&Distance	&M$_{\star}$	&$v_{sys}$	&$i_{jet}$	&PA$_{jet}$ &References	\\
		&		&(pc)		&(M${_\odot}$)	&(km\,s$^{-1}$)	&(deg)	&(deg)	&		\\ 
\tableline
DG\,Tau		&Taurus		&140		&0.67		&+16.5		&52		&226 &1, 2		\\
CW\,Tau		&Taurus		&140		&1.4		&+14.5		&41		&155 &3, 4, 5	\\
HH\,30		&Taurus		&140		&0.45		&+21.5		&1		&33 &6, 7, 8		\\
TH\,28		&Lupus 3	&170		&...		&+5		&10		&98 &9, 10		\\ 
\tableline
\end{tabular}}
\end{center}
\caption{Details of T\,Tauri jet targets investigated in this paper. All radial velocity results, $v_{rad}$, are quoted after correction for the systemic heliocentric radial velocity, $v_{sys}$. The inclination angle of the jet, $i_{jet}$, is 
given with respect to the plane of the sky. Values for the jet position angle, PA$_{jet}$, were determined from archival {\em HST} images. For our observations, we requested a slit position angle of 90$^{\circ}$ with respect to PA$_{jet}$, Figure\,\protect\ref{orientation}. References - (1) \protect\protect\citealp{Eisloffel98}; (2) \protect\protect\citealp{Bacciotti02}; (3) \protect\protect\citealp{GomezdeCastro93}; 
(4) \protect\protect\citealp{Hartmann86}; (5) \protect\protect\citealp{Hartigan04}; (6) \protect\protect\citealp{Pety06}; (7) \protect\protect\citealp{Appenzeller05}; 
(8) \protect\protect\citealp{Mundt90}; (9) \protect\protect\citealp{Graham88}; (10) \protect\protect\citealp{Krautter86}. 
\label{targets}}
\end{table*}

{\em HST}/STIS optical observations were conducted for the approaching jets from T\,Tauri stars DG\,Tau, CW\,Tau and HH\,30 
(proposal ID 9435 -- observations of the other jets in the same proposal are described in \protect\citealp{Coffey04}). The three 
jets were observed with slit offsets of 0$\farcs$3, 0$\farcs$2 and 0$\farcs$6 respectively. These correspond to a 
deprojected distance along each jet of 68\,AU, 37\,AU and 84\,AU. The larger offset for HH\,30 was deemed more appropriate since the 
star and the jet base are obscured by the system's edge-on disk. The CCD detector was used with the G750M grating, 
centred on 6581\,\AA, and a slit of aperture 52$\times$0.1\,arcsec${^2}$. Spectral sampling was 0.554\,\AA\,pixel$^{-1}$, 
corresponding to a radial velocity sampling of $\sim$25\,km\,s$^{-1}$\,pixel$^{-1}$, and spatial sampling was 0.$\arcsec$05\,pixel$^{-1}$. The effective spectral resolution, achieved using Gaussian fitting, is typically one fifth of the sampling, while the spatial resolution is twice the sampling (i.e. 2 pixel resolution of 0\farcs1). 
One long exposure of $\sim$2100\,s was made of the approaching jet from DG\,Tau on Dec~1~2003 and the approaching jet from CW\,Tau 
on Dec~21~2003, while two long exposures (to be co-added) were made of the approaching jet from HH\,30 on Nov~30~2003. 
The data were calibrated through the standard {\em HST} pipeline, subtraction of any reflected stellar continuum was performed, and signals from defective pixels were removed. 

{\em HST}/STIS NUV observations were conducted for the approaching jet from DG\,Tau, and each lobe of the bipolar jet from TH\,28 
(proposal ID 9807). A slit offset of 0.$\arcsec$3 was used in each case. For TH\,28, this corresponds to a deprojected 
distance along the jet of 52\,AU. The NUV MAMA detector was used with the E230M echelle grating, centred on 2707\,\AA, and a long 
slit of 
aperture 6$\times$0.2\,arcsec${^2}$ to ensure the full width of the jet was observed. Spectral sampling was 
0.045\,\AA\,pixel$^{-1}$, corresponding to a radial velocity sampling of $\sim$5\,km\,s$^{-1}$\,pixel$^{-1}$, and spatial sampling was 
0.$\arcsec$029\,pixel$^{-1}$. The effective spectral resolution achieved was on average 5\,km\,s$^{-1}$, while the spatial resolution was twice the sampling (i.e. 2 pixel resolution of 0\farcs058). Two long exposures of $\sim$2500\,s (to be co-added) were made of the approaching jet of 
DG\,Tau on Dec~1~2003, and each lobe of the bipolar jet from TH\,28 on Jun~18~2004. Although the data were largely 
processed through the standard {\em HST} pipeline, the combination of an echelle grating with a long slit meant that wavelength 
calibration was not conducted as part of the pipeline procedure. Wavelength calibration was therefore necessary in the initial stages of data reduction, and was carried out using standard IRAF routines. 

\section{Results}
\label{results}

In our chosen observing mode, the slit does not cover the star. The instrument spatial line spread function has a half-width at zero-maximum of 0\farcs15 in the optical regime and 0\farcs145 in the NUV. Therefore, observed emission lines in both the 
optical and NUV regions originate in jet material excited through low-velocity shocks 
\protect\cite{Hartigan87}. All optical targets were found to emit in the forbidden emission line doublets [\ion{O}{1}], [\ion{N}{2}] and [\ion{S}{2}] (i.e. at vacuum rest wavelengths of 6302.046,\,6365.537\,\AA,\, 6549.86,\,6585.27\,\AA\, and 6718.29,\,6732.67\,\AA\, respectively). All three jet targets observed in the NUV were found to be strong emitters in both lines of the \ion{Mg}{2} doublet (i.e. at 
vacuum rest wavelengths of 2796.352\,\AA\, and 2803.531\,\AA). Their profiles are broad, extending over some 200\,km\,s$^{-1}$, 
consistent with the idea of shock excitation. For each target, the extension of the broad line profile in the NUV is approximately the same as for the optical lines presented in this and previous studies but, spatially, the emission region appears to be more concentrated on the jet axis. In the NUV, spatial extention is represented by FWHM values in the range of 0\farcs12 to 0\farcs20 for 
both jet lobes from TH\,28 (as opposed to 0\farcs27 and 0\farcs41 for [\ion{O}{1}] and [\ion{S}{2}] receding jet lines respectively -- from 
optical spectra taken with {\em HST}/STIS, proposal ID 9435). Values for the approaching jet from DG\,Tau are similar, with FWHM values of 0\farcs12 in the NUV (as opposed to 
0\farcs15 and 0\farcs20 for [\ion{O}{1}] and [\ion{S}{2}] optical lines respectively). The trend shows that the high excitation NUV lines preferentially originate closer to the axial region of the jet. This is consistent with the expectations of jet models with shock heating, according to which layers closer to the axis are more excited than those further away, since the velocity of the shock front is higher on-axis and gives rise to higher shock temperatures there. 

\subsection{Qualitative analysis}

If a jet is rotating, we would expect to observe a difference in the magnitude of the Doppler shift between material on one side and the other of a jet's symmetry axis (see \protect\citealp{Coffey04}). Therefore, signatures of jet rotation are expected to present themselves in the form of a contour tilt in position-velocity diagrams of jet spectral emission lines. In other words, the tilt in contours illustrates that radial velocities are lower on one side of the jet, thus giving a qualitative indication of systematic radial velocity differences as expected from a rotating jet. 

\begin{figure*}
\begin{center}
\includegraphics[scale=0.3]{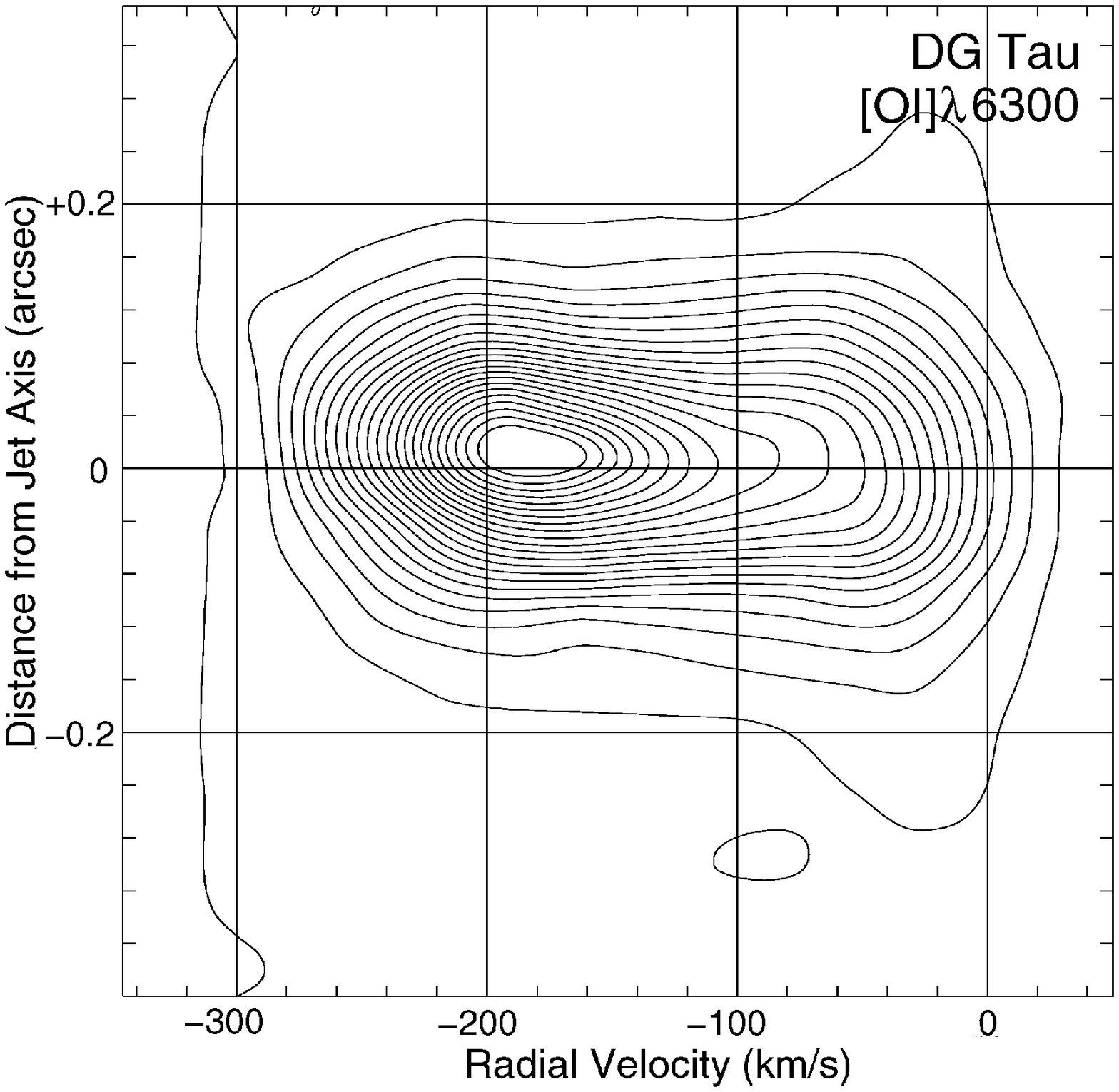}\includegraphics[scale=0.3]{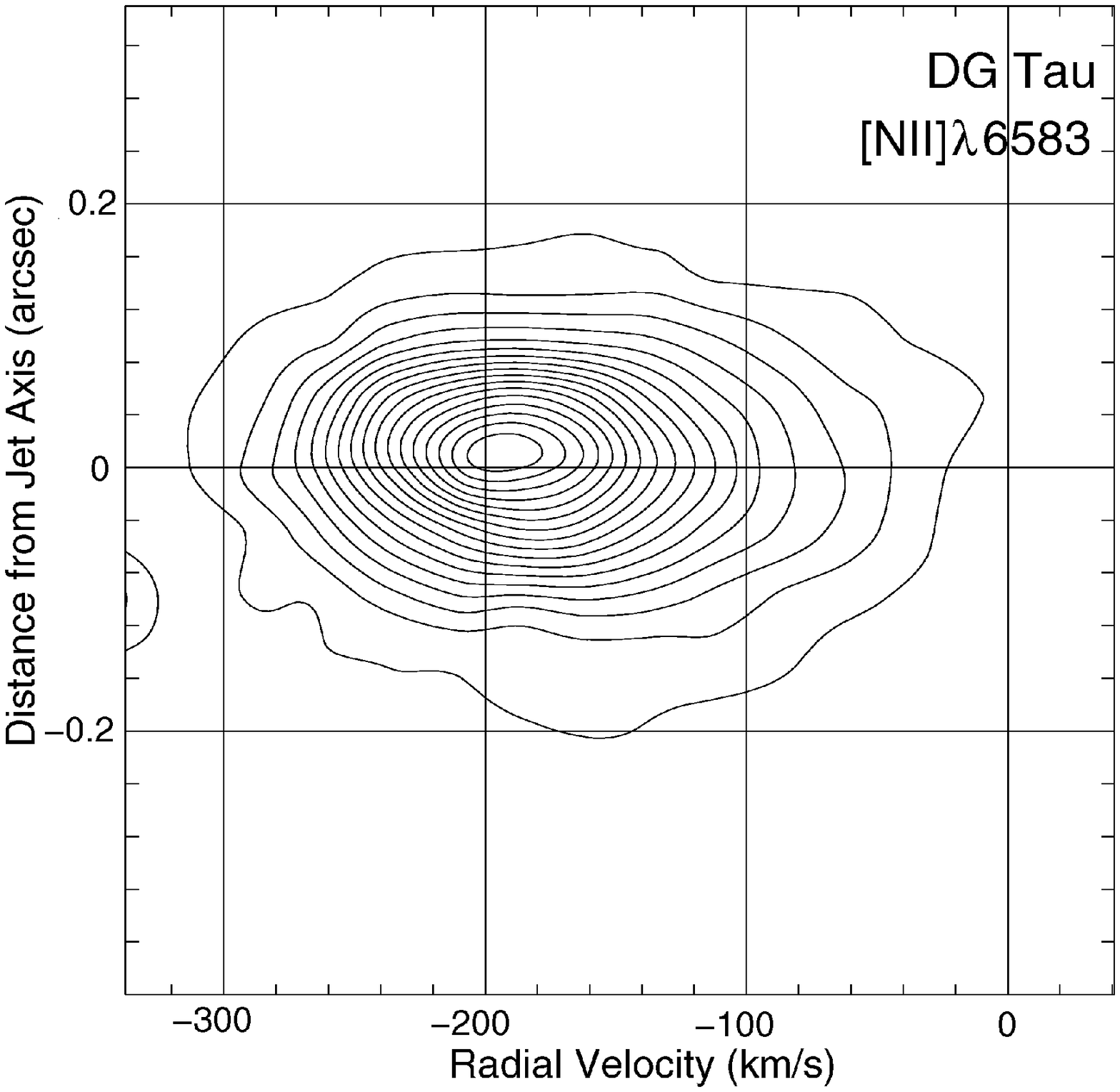}\includegraphics[scale=0.3]{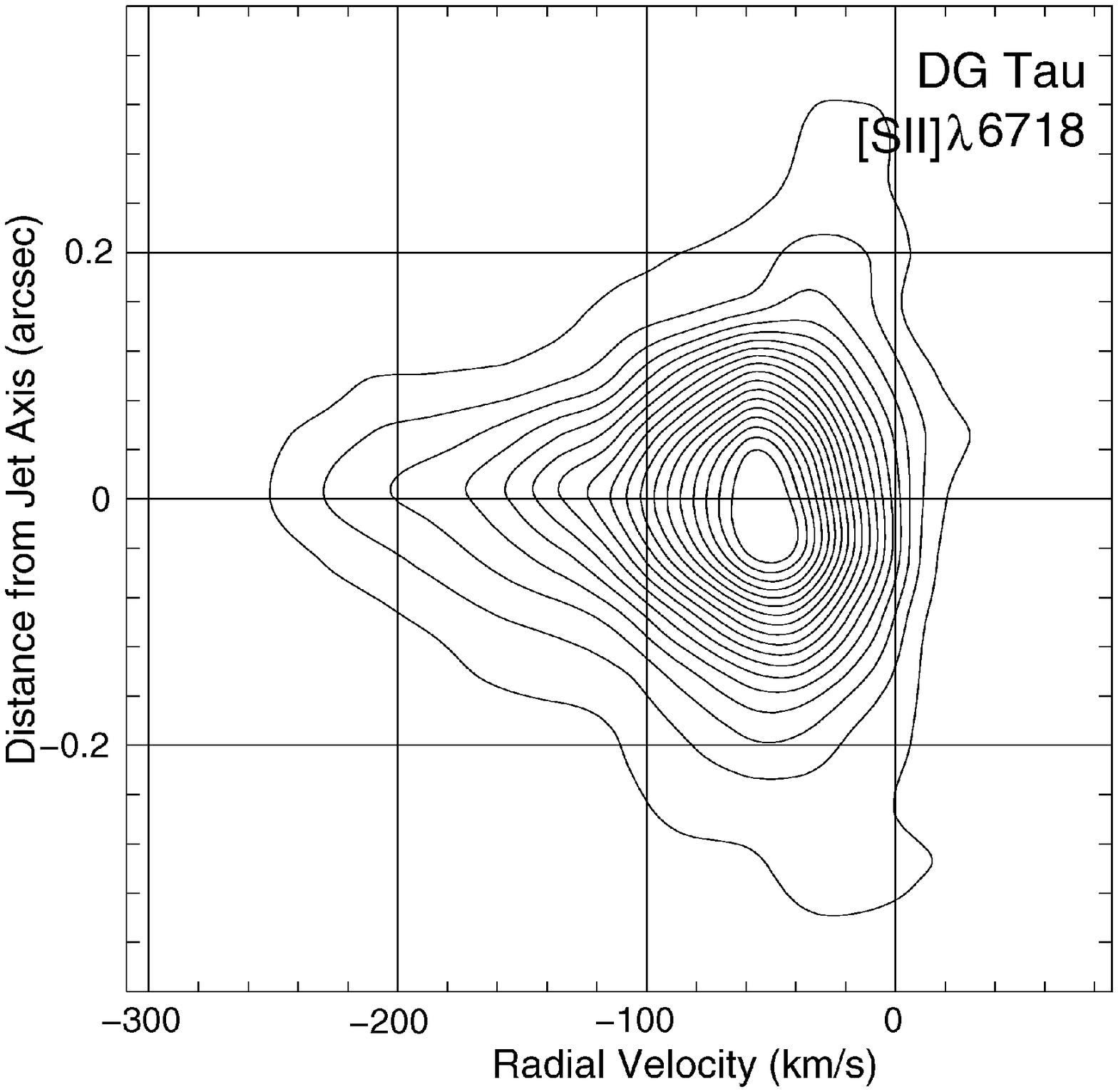}\\
\hspace{0.1 cm}\includegraphics[scale=0.3]{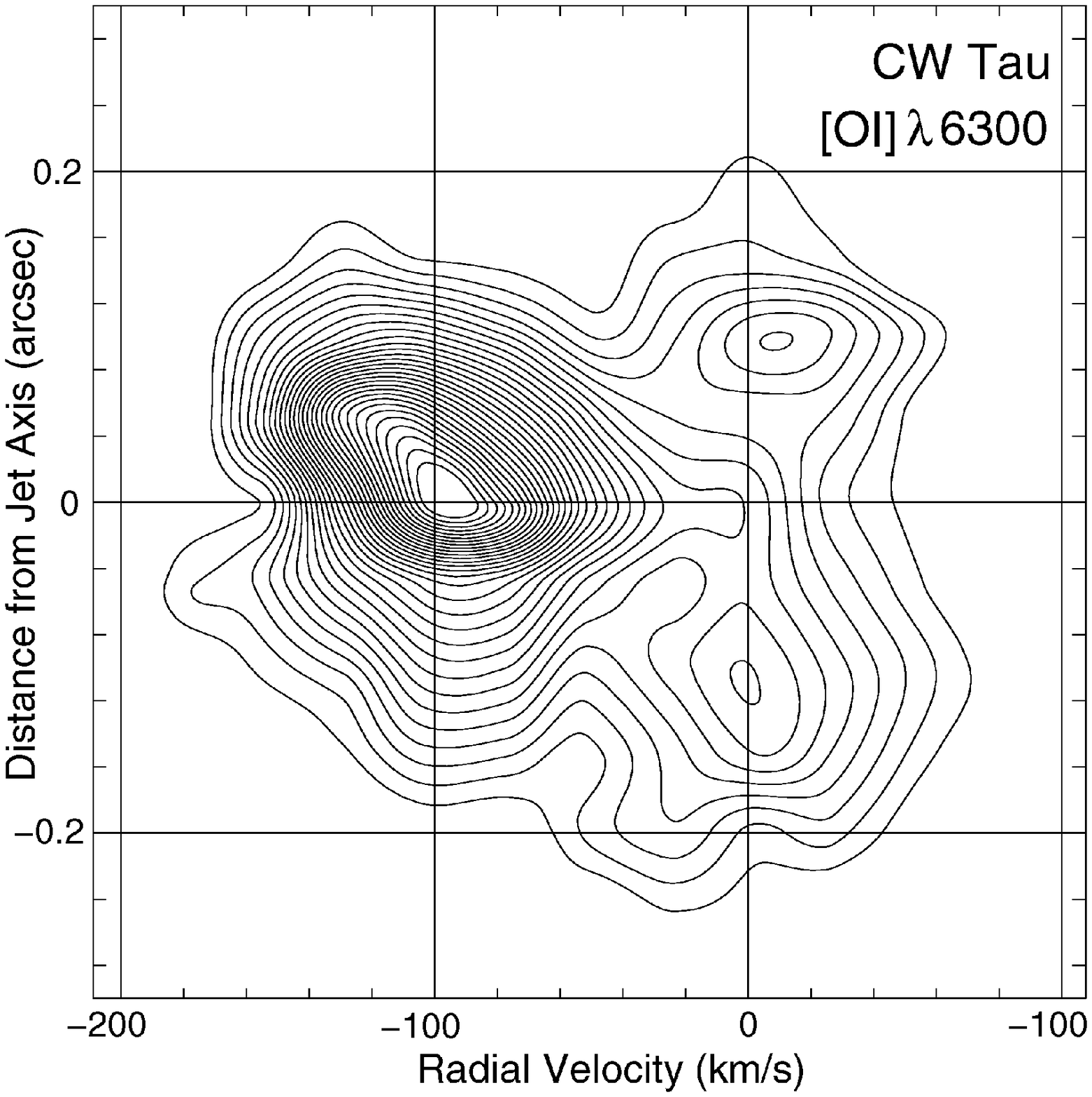}\includegraphics[scale=0.3]{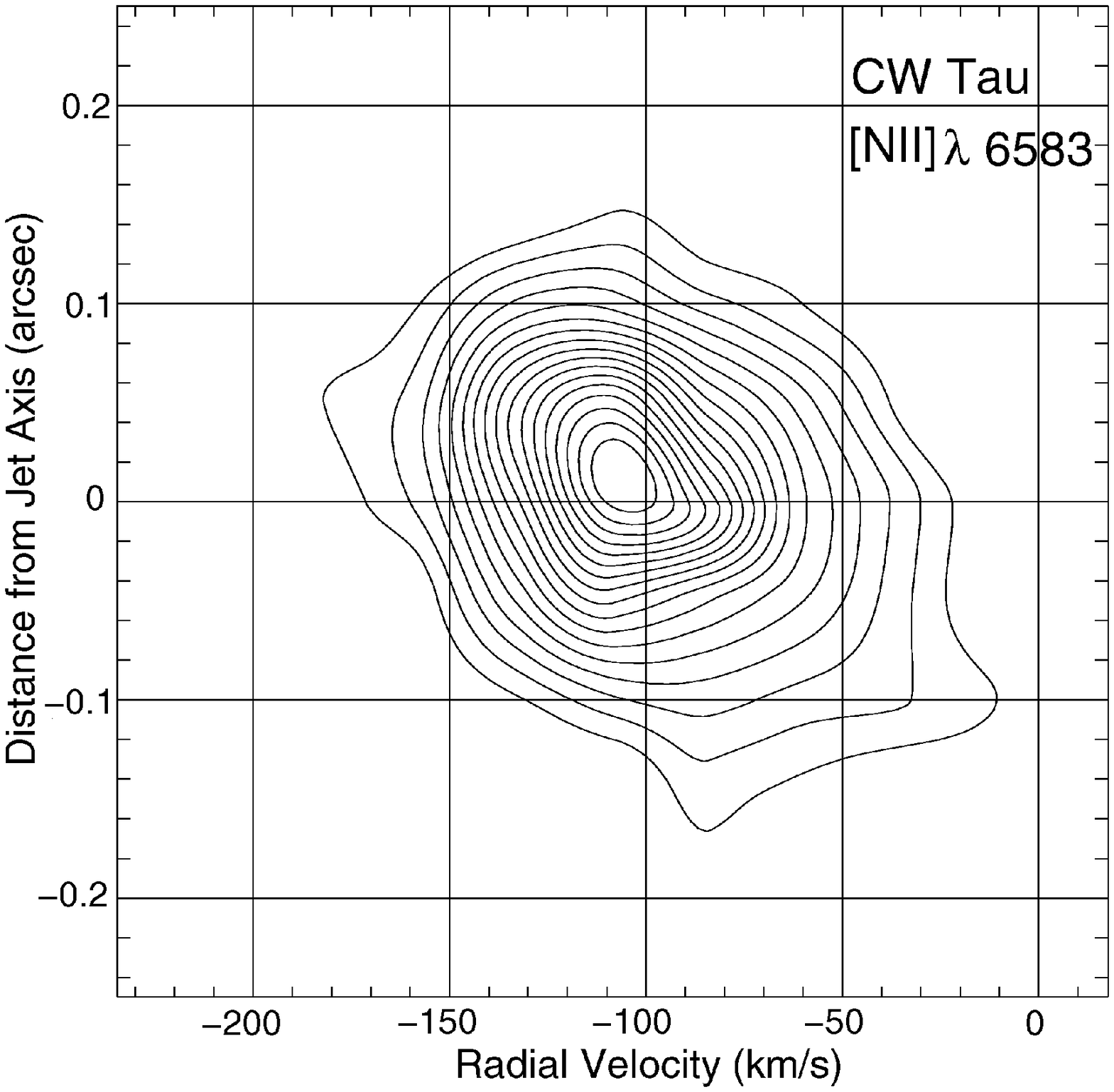}\includegraphics[scale=0.3]{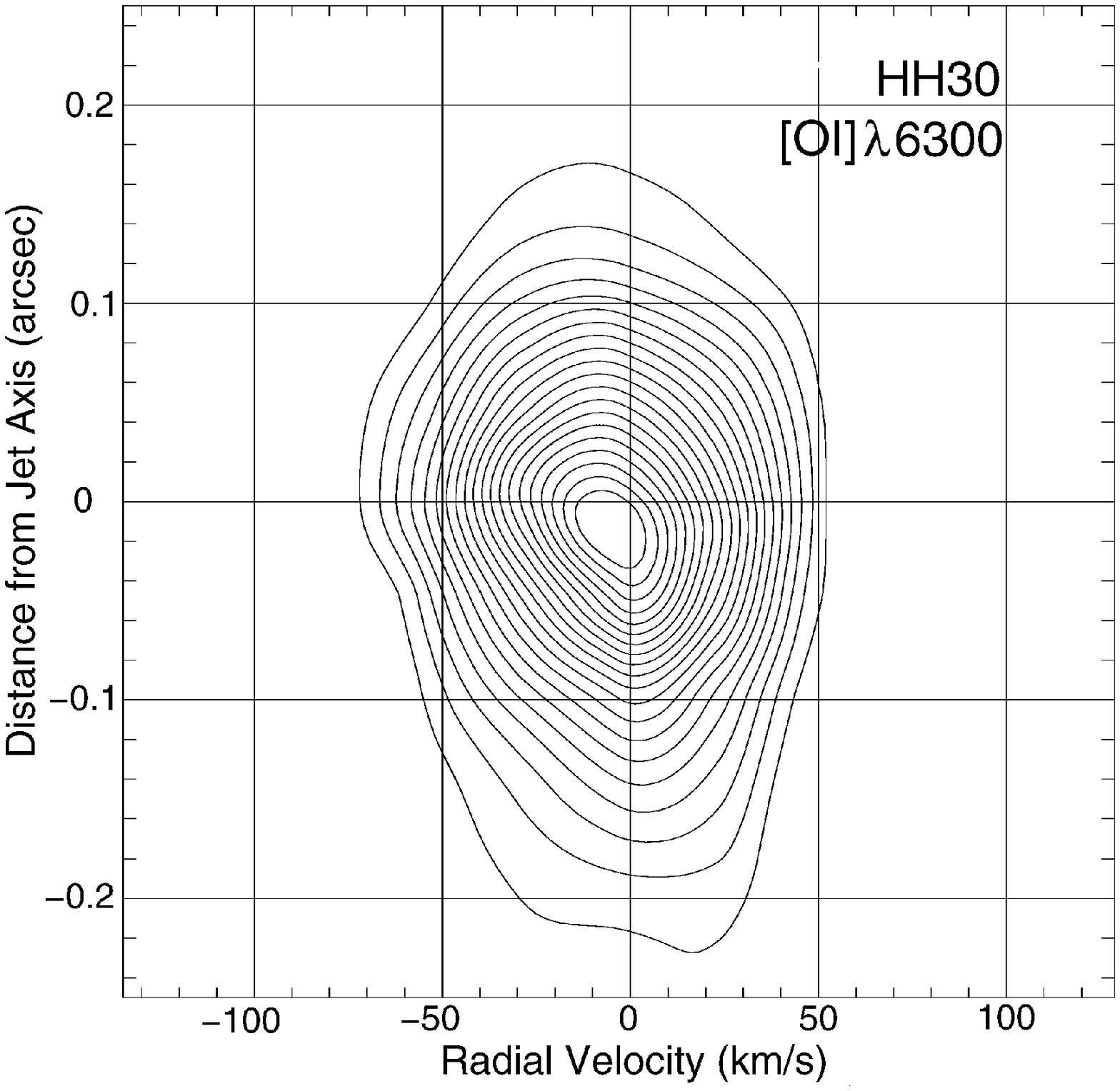}
\figcaption{Position-velocity contour diagrams in selected emission lines for each jet targets observed in the optical regime. Note that the [\ion{O}{1}] emission traces both the higher and lower velocity jet material, while the [\ion{N}{2}] emission is only seen at high velocities and the [\ion{S}{2}] emission is mainly concentrated at low velocities. Furthermore, higher velocity emission appears to be more concentrated on the jet axis, while lower velocity emission is spatially extended. The skew in contours is indicative of jet rotation. 
The positive direction of the y-axis is illustrated 
pictorially as the slit direction in Figure\,\protect\ref{orientation}. 
Plots are corrected for systemic heliocentric radial 
velocity, $v_{sys}$, Table\,\protect\ref{targets}. Contour values of the fluxes for each panel, in units of erg\,cm$^{-2}\,$sec$^{-1}$\,\AA$^{-1}$\,arcsec$^{-2}$, are given in Table\,\protect\ref{contour_intervals}. 
\label{pv_optical_jets}}
\end{center}
\end{figure*}

\begin{figure*}
\begin{center}
\includegraphics[scale=0.35]{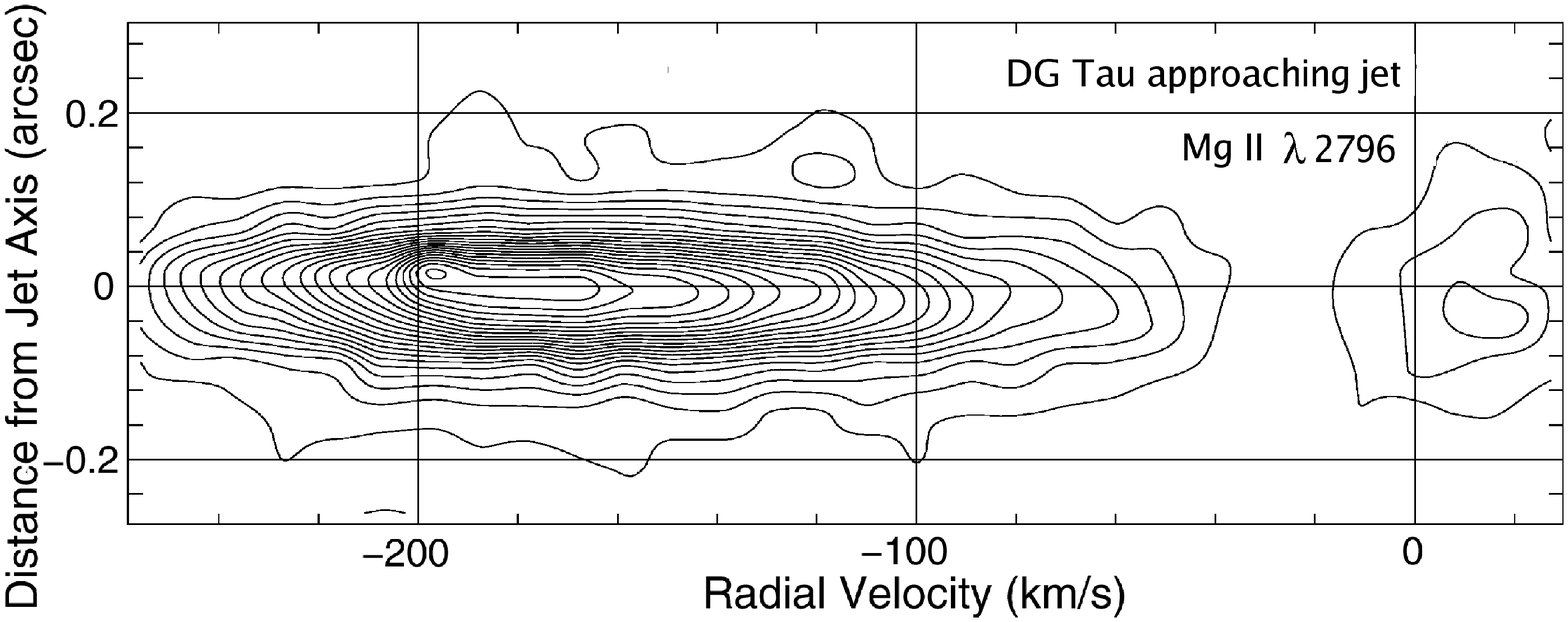}
\includegraphics[scale=0.35]{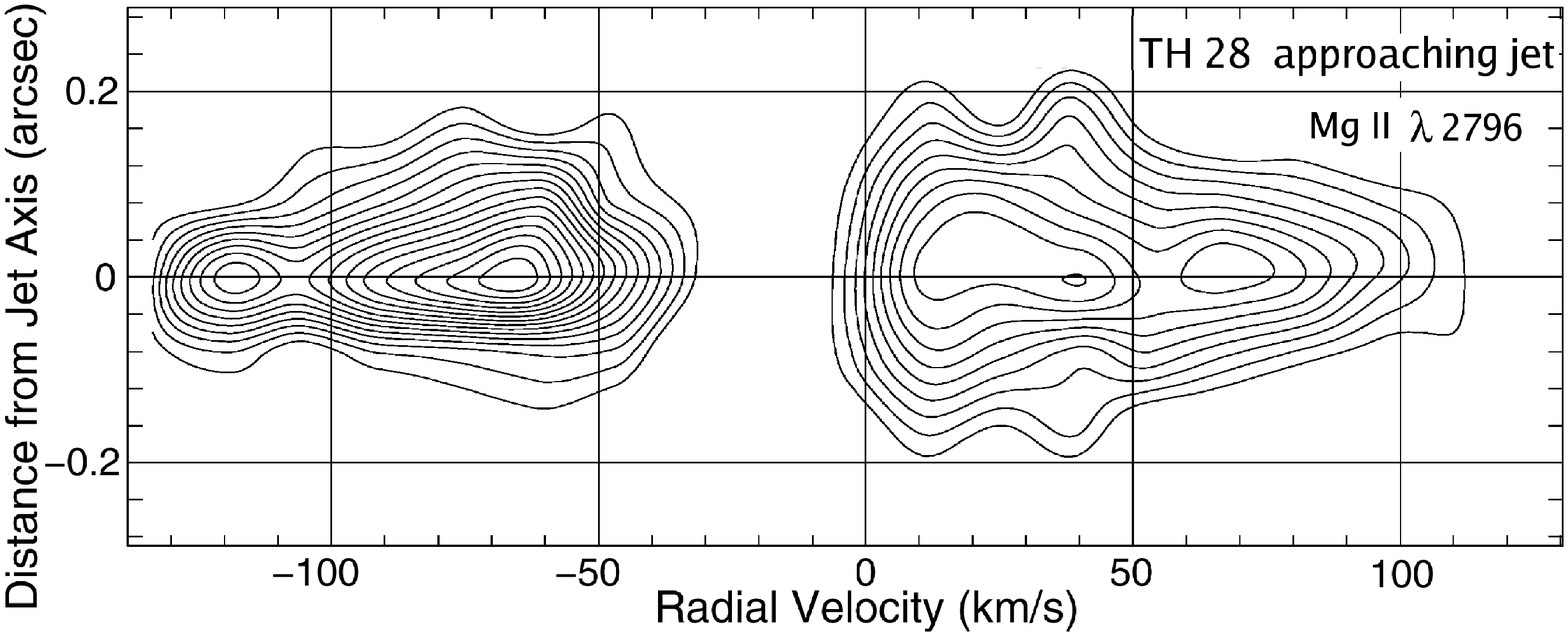}
\includegraphics[scale=0.35]{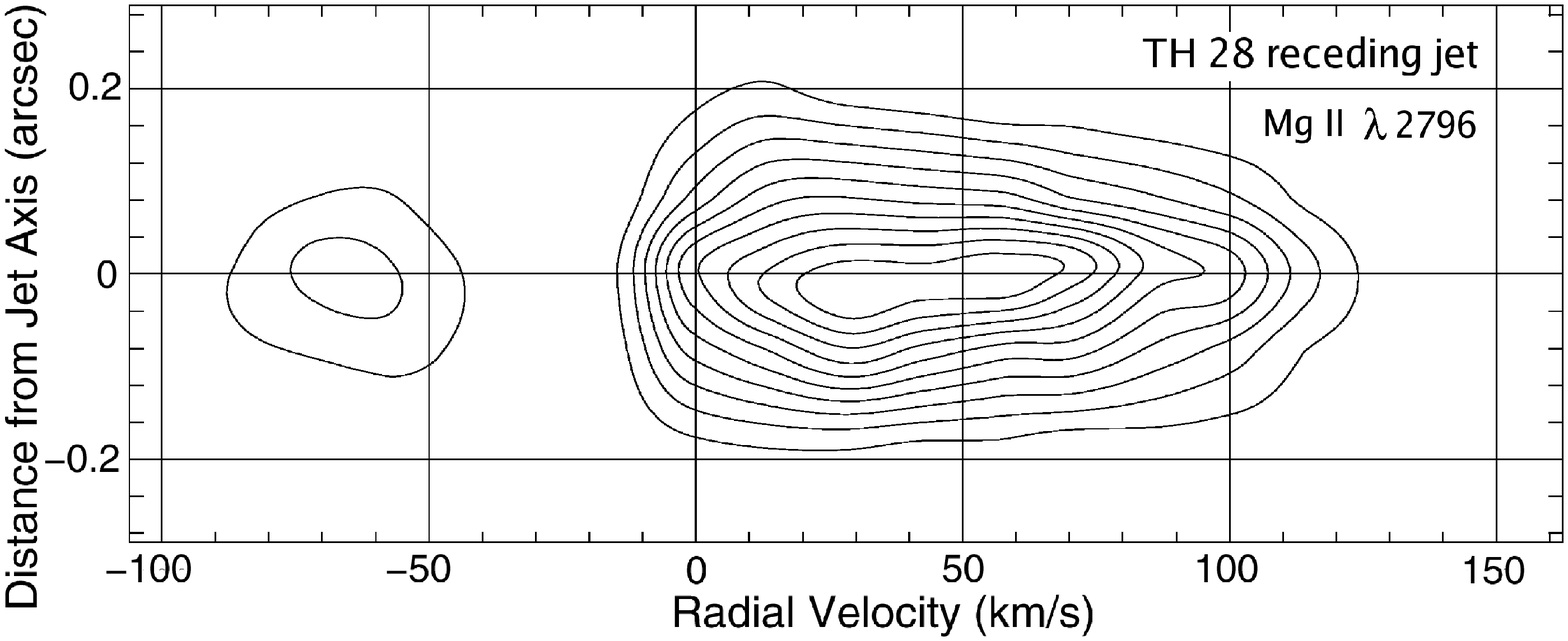}
\figcaption{Position-velocity contour diagrams in \ion{Mg}{2}\,$\lambda$2796 emission for the jet targets observed in the NUV regime. The positive direction of the y-axis is illustrated pictorially as the slit direction in Figure\,\protect\ref{orientation}. 
Plots are corrected for systemic heliocentric radial 
velocity, $v_{sys}$, Table\,\protect\ref{targets}. Contour values of the fluxes for each panel, in units of erg\,cm$^{-2}\,$sec$^{-1}$\,\AA$^{-1}$\,arcsec$^{-2}$, are given in Table\,\protect\ref{contour_intervals}. 
\label{pv_nuv_jets}}
\end{center}
\end{figure*}

Figures\,\protect\ref{pv_optical_jets} and \protect\ref{pv_nuv_jets} show position-velocity contour diagrams for selected optical and NUV emission lines, the contour levels for which are given in Table\,\protect\ref{contour_intervals}. In each contour plot, the positive direction of the y-axis corresponds to the slit direction as illustrated in Figure\,\protect\ref{orientation}. While a position-velocity contour tilt can generally be recognised at optical wavelengths, a tilt is not so apparent in the NUV range. It is likely that the broad profile shape in the dispersion direction combined with the relatively narrow spatial FWHM may detract from a visible delineation of any trend. The existence of an absorption feature for the NUV permitted transitions may also make any subtle contour trend less evident. Nevertheless, quantitative measurements yielded positive results. 

\begin{table*}
\begin{center}
\scriptsize{\begin{tabular}{lcccc}
\hline \hline
Target			&Emission		&Contour	&Contour	&Contour	
\\
			&line			&Floor		&Ceiling	&Interval	
\\
\hline
DG\,Tau approaching jet	& [O I]]$\lambda$6300 &1.0$\times$10$^{-15}$ &4.4$\times$10$^{-13}$ &2.0$\times$10$^{-14}$ \\
			&[N II]$\lambda$6583 &5.0$\times$10$^{-15}$ &8.1$\times$10$^{-14}$ &4.0$\times$10$^{-15}$ \\
			&[S II]$\lambda$6716 &3.0$\times$10$^{-15}$ &7.1$\times$10$^{-13}$ &4.0$\times$10$^{-15}$ \\
			&Mg\,II\,$\lambda$2796 &1.0$\times$10$^{-13}$ &2.1$\times$10$^{-12}$ &1.0$\times$10$^{-13}$ \\
CW\,Tau approaching jet	& [O I]$\lambda$6300 &3.0$\times$10$^{-15}$ &3.2$\times$10$^{-14}$ &1.0$\times$10$^{-15}$ \\
			&[N II]$\lambda$6583 &2.0$\times$10$^{-15}$ &1.3$\times$10$^{-14}$ &7.0$\times$10$^{-16}$ \\
HH\,30 approaching jet 	& [O I]$\lambda$6300 &7.0$\times$10$^{-15}$ &6.4$\times$10$^{-14}$ &3.0$\times$10$^{-15}$ \\ 
TH\,28 approaching jet 	&Mg\,II\,$\lambda$2796 &1.0$\times$10$^{-13}$ &3.9$\times$10$^{-13}$ &2.1$\times$10$^{-14}$ \\
TH\,28 receding jet 	&Mg\,II\,$\lambda$2796 &2.0$\times$10$^{-13}$ &2.0$\times$10$^{-12}$ &1.3$\times$10$^{-13}$ \\\hline
\end{tabular}}
\end{center}
\caption{Contour intervals of the emission line surface brightness of position-velocity diagrams, 
Figure\,\protect\ref{pv_optical_jets}, in units of erg\,cm$^{-2}$\,s$^{-1}$\,\AA$^{-1}$\,arcsec$^{-2}$. 
\label{contour_intervals}}
\end{table*}

\subsection{Quantitative analysis}
\label{Quantitative_analysis}

Quantitative measurements of jet rotation require determination of differences in radial velocity at positions {\em equidistant} either side of the jet axis. The intensity peak in the spatial direction of the jet emission is assumed to mark the physical jet axis. The physical jet axis must be aligned with the centre of a pixel in order to measure  {\em equidistant} either side of it. In practice, emission peaks were offset slightly (in most cases by about 0.25 pixels from the central pixel row) due to offsets in telescope and instrument pointing. Therefore, the first step in the analysis was recentre the emission. The method was to perform a Gaussian fit on each column (i.e in the spatial direction), identify the column with the highest amplitude fit, determine the offset of its peak, and recentre the 2D emission according to this offset amount using an interpolation technique. The precision reached was $\pm$\,0$\farcs$01 (i.e. one fifth of the spatial sampling of 0$\farcs$05). Offsets were double-checked by binning the emission onto the spatial axis and Gaussian fitting the peak. The offsets were in agreement for the optical data, but only the latter technique was relied upon when processing the NUV data. Note that the position-velocity diagrams (Figures\,\ref{pv_optical_jets} and \ref{pv_nuv_jets}) are shown prior to this recentering, specifically to illustrate the presence of contour tilts before the data were processed in any way. 

\begin{figure*}
\begin{center}
\includegraphics[scale=0.8]{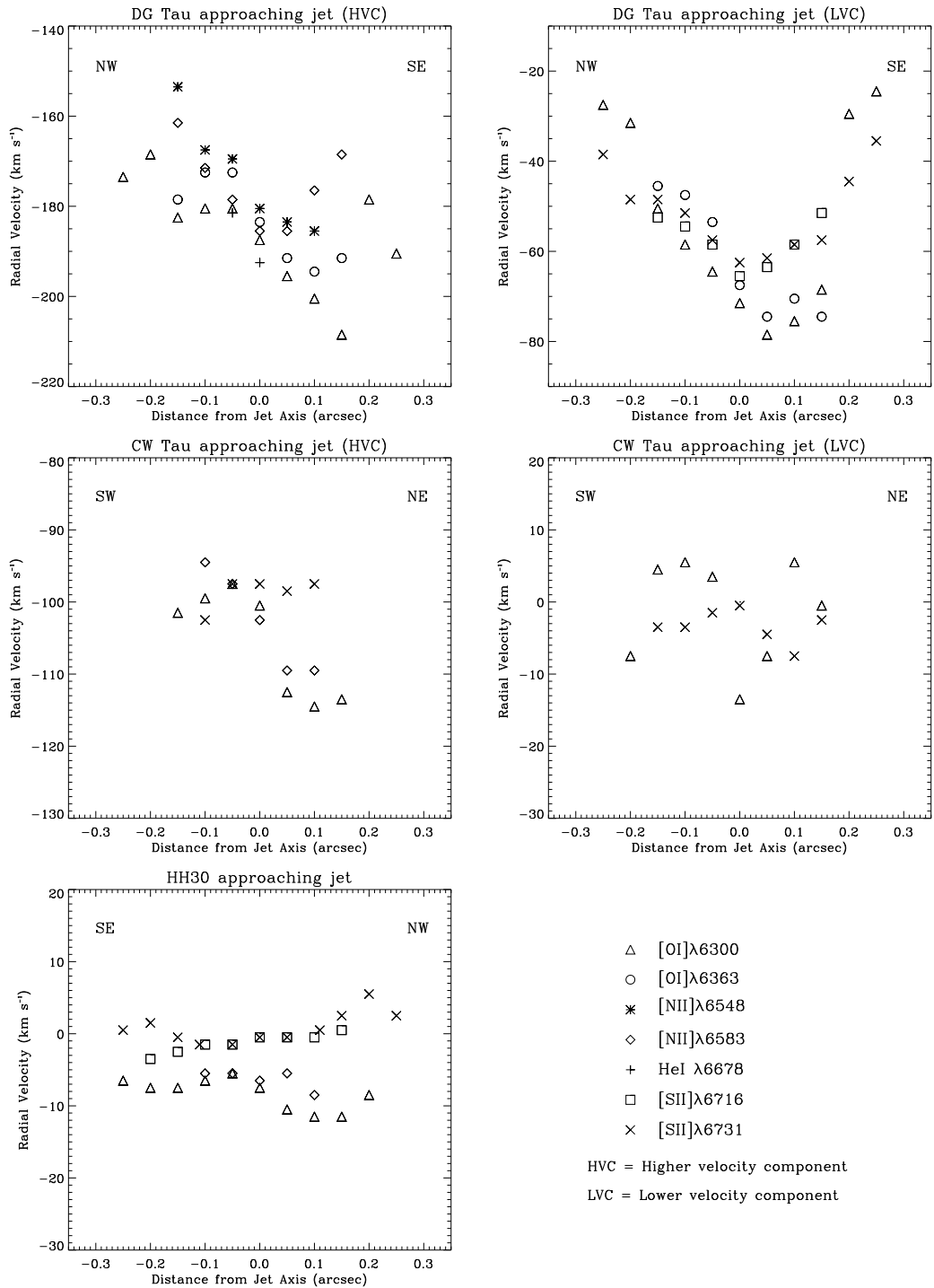}
\figcaption{Radial velocity profiles across the jets in various optical emission lines. All radial velocities are corrected for systemic heliocentric radial velocity. Profile asymmetries were interpreted as indications of jet rotation. 
\label{velocityprofiles_opt}}
\end{center}
\end{figure*}

\begin{figure*}
\begin{center}
\includegraphics[scale=1.15]{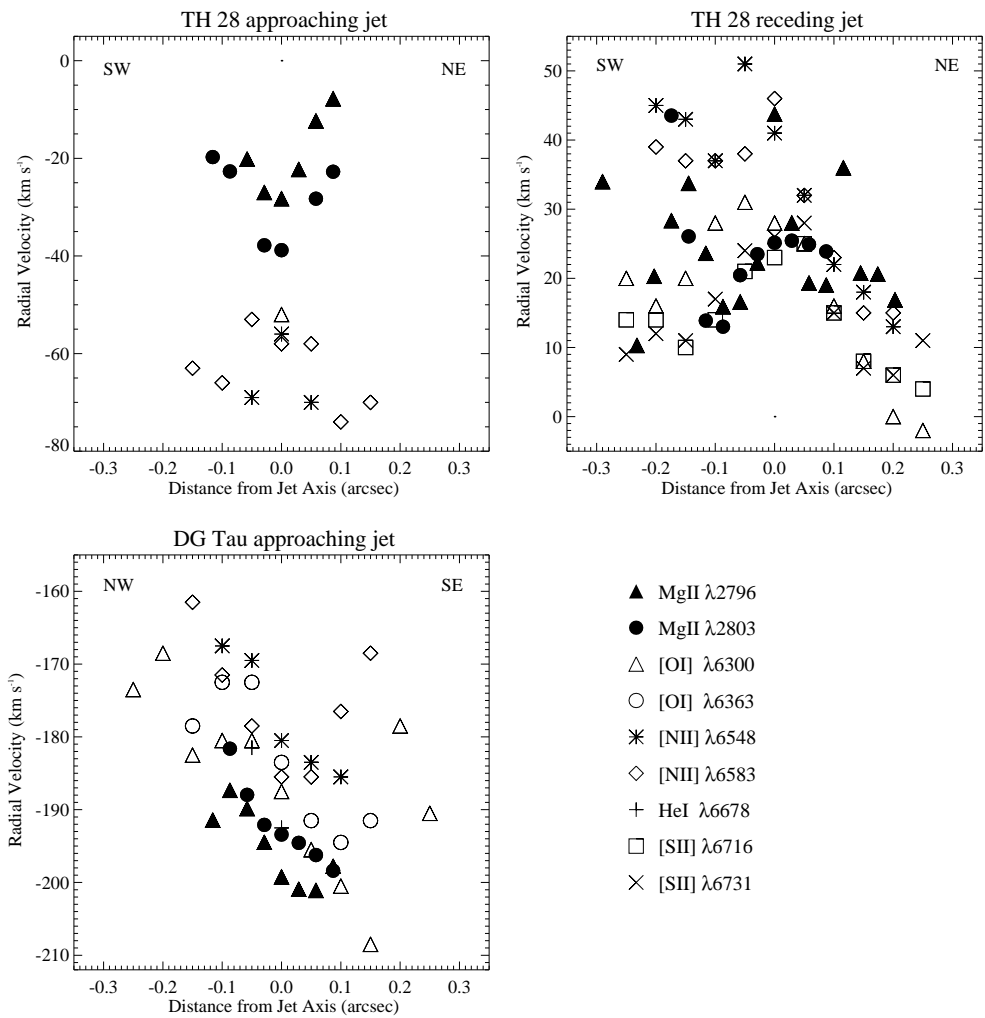}
\figcaption{Radial velocity profiles across the jets in 
NUV emission lines (similar to Figure\,\ref{velocityprofiles_opt}). The optical 
datasets are overlaid on the NUV datasets, to 
illustrate how well the results for the two 
wavelength regions fit together. Optical data for TH\,28 were published previously \protect\cite{Coffey04}. The NUV profiles for the TH\,28 bipolar jet should be taken as an indication only, because Gaussian fitting proved very difficult and so an accurate profile could not be constructed. The radial velocity differences in Table\,\ref{rotational_velocities_uv} were obtained by relying on a cross-correlation routine, as described in Section\,\ref{Quantitative_analysis}.
\label{velocityprofiles_optuv}}
\end{center}
\end{figure*}

Gaussian fitting was then used to determine where the emission peaked in each pixel row parallel to the jet axis (i.e. in the dispersion direction). From this analysis, radial velocities were determined at each distance from the axis and a radial velocity profile of the jet in the transverse direction was obtained. Radial velocity profiles are plotted in Figures\,\ref{velocityprofiles_opt} and \ref{velocityprofiles_optuv} (in which the orientation, derived from Figure\,\ref{orientation}, is given in the upper corners of each plot). Where the low and high velocity jet material is clearly distinguishable, the plots have been divided in to low and high velocity components. Plots reveal that the on-axis jet material is travelling at the highest velocity, while the borders of the jet are travelling at lower velocities. However, the velocity profiles are not symmetrical. Velocities on one side of the axis are lower than on the other. Furthermore, each emission line traces a different degree of asymmetry. In general, [\ion{O}{1}] lines are {\em clear} tracers of radial velocity asymmetry while [\ion{S}{2}] lines show much more symmetric profiles. This combination of contributions from different emission lines and different velocity components can draw from the clarity of some of these profile plots. The clearest case is the plot for the TH\,28 receding jet at optical wavelengths. 

\begin{figure*}
\begin{center}
\includegraphics[scale=0.8]{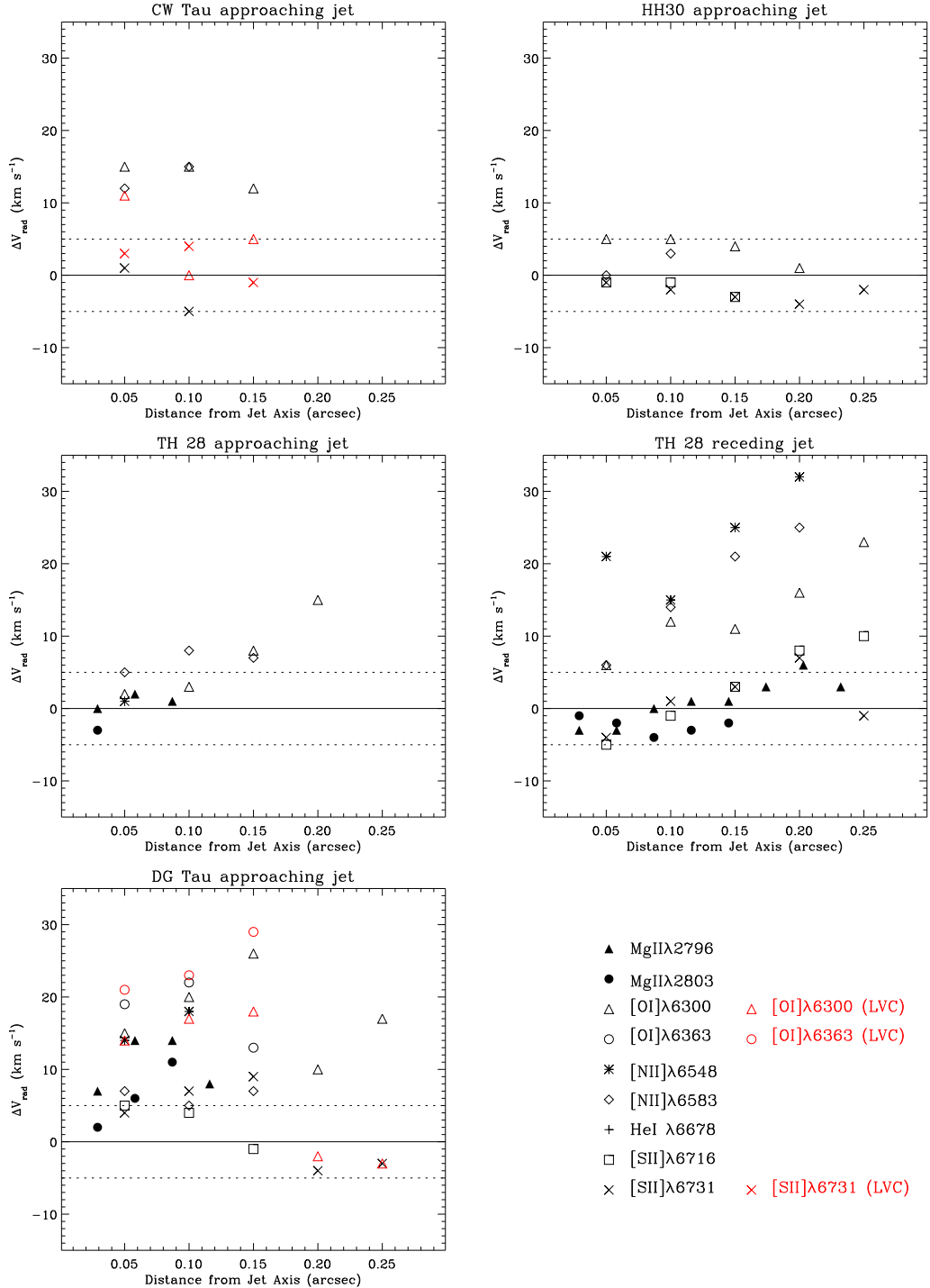}
\figcaption{Transverse radial velocity differences measured in optical and NUV lines, for a slit placed across the jet less than 100\,AU 
from the star. These values are taken from Tables\,\ref{rotational_velocities_optical} and 
\ref{rotational_velocities_uv}, with the exception of those for the optical lines in the TH\,28 receding jet, which are 
taken from Coffey~et~al.~(2004). The horizontal strip marks the maximum error bar in the velocity difference about zero, but is more representive of the typical error on NUV data points than optical data points (see Figure\,\ref{errorbars_opt_nuv}). 
\label{velocity_shifts}}
\end{center}
\end{figure*}

The differences in radial velocities were cross-checked by comparing them to the results of a cross-correlation routine. While the latter has the advantage of being independent of the line profile shape, it was found to be very sensitive to the extent and level of the background. Although there is generally good agreement between the methods (within 1 or 2\,km\,s$^{-1}$), Gaussian fitting proved more reliable in the optical region where all emission lines were of suitable profile shape. In the NUV range, in the case of DG\,Tau where the higher velocity material escaped absorption, Gaussian fitting proved suitable and was found to be in reasonable agreement with cross-correlation (within 2\,km\,s$^{-1}$). In the remaining cases, Gaussian fitting was not reliable due to absorption of part of the profile. In these cases, measurement of radial velocity differences relied on cross-correlation. Tables\,\ref{rotational_velocities_optical} and \ref{rotational_velocities_uv} give radial velocity differences, $\Delta 
v_{rad}$, for individual emission lines in each jet lobe. These measurements are also illustrated graphically in Figure\,\ref{velocity_shifts}. Asymmetries were typically 10 to 25\,km\,s$^{-1}$ in the optical region, and 10\,km\,s$^{-1}$ in the NUV region. 

\begin{figure*}
\begin{center}
\includegraphics[scale=0.925]{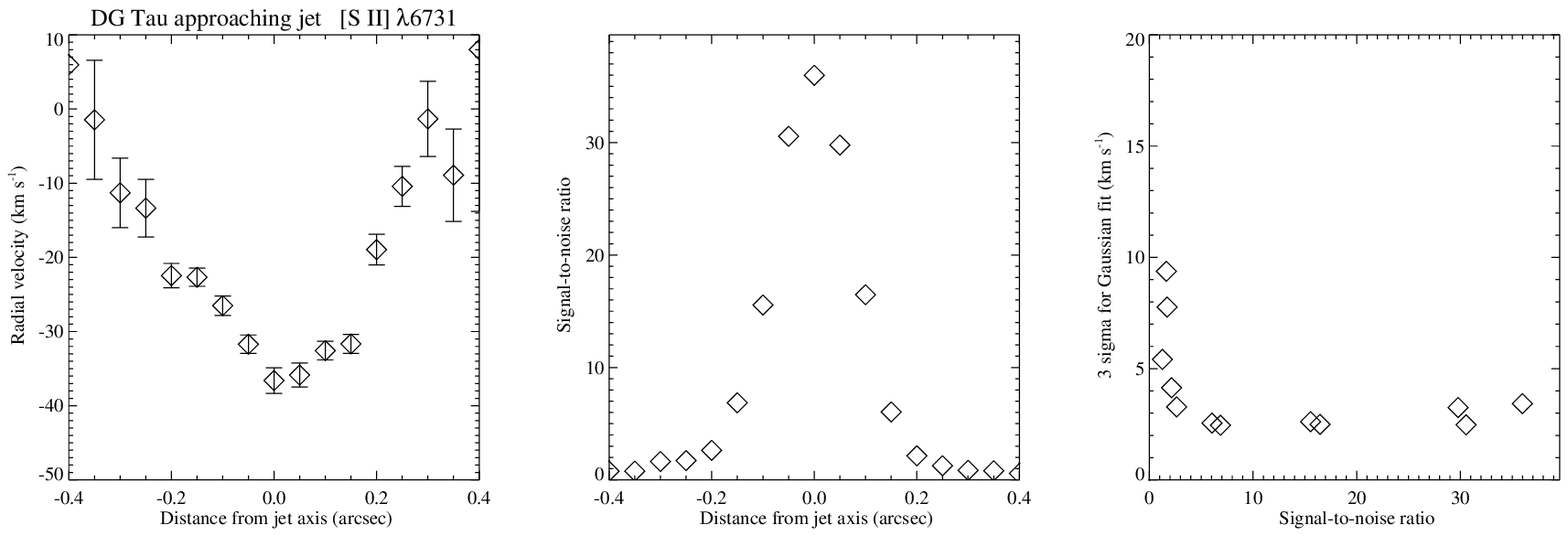}
\includegraphics[scale=0.925]{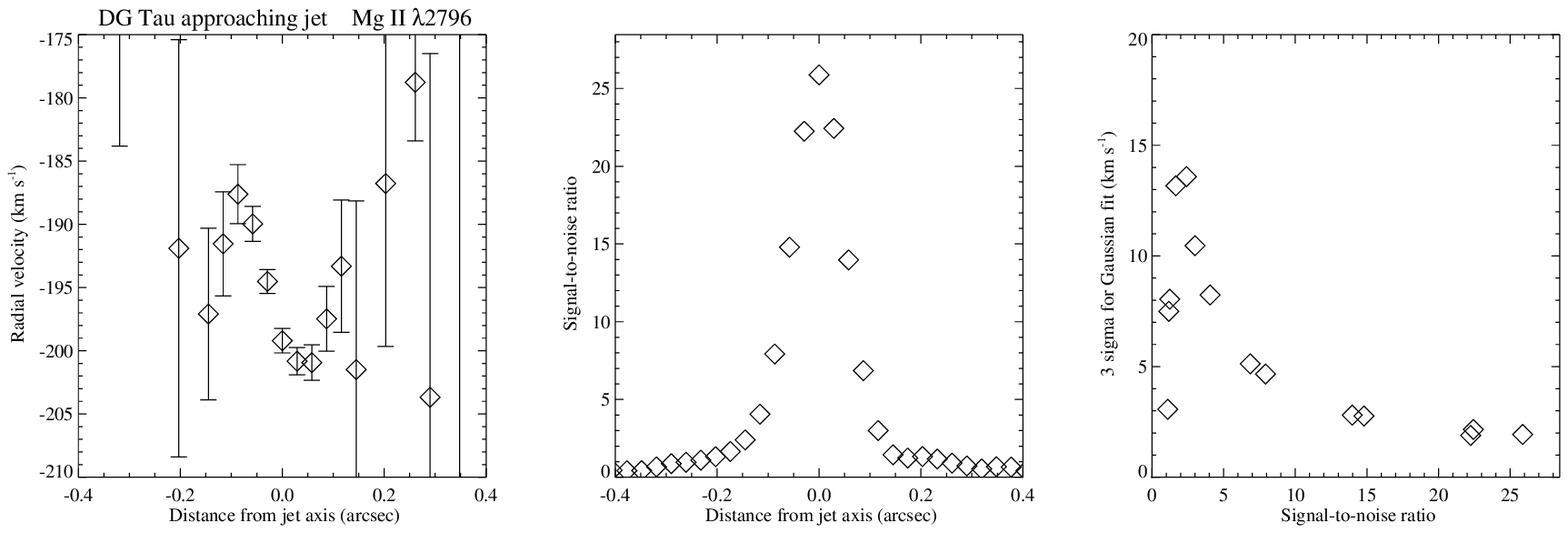}
\includegraphics[scale=0.925]{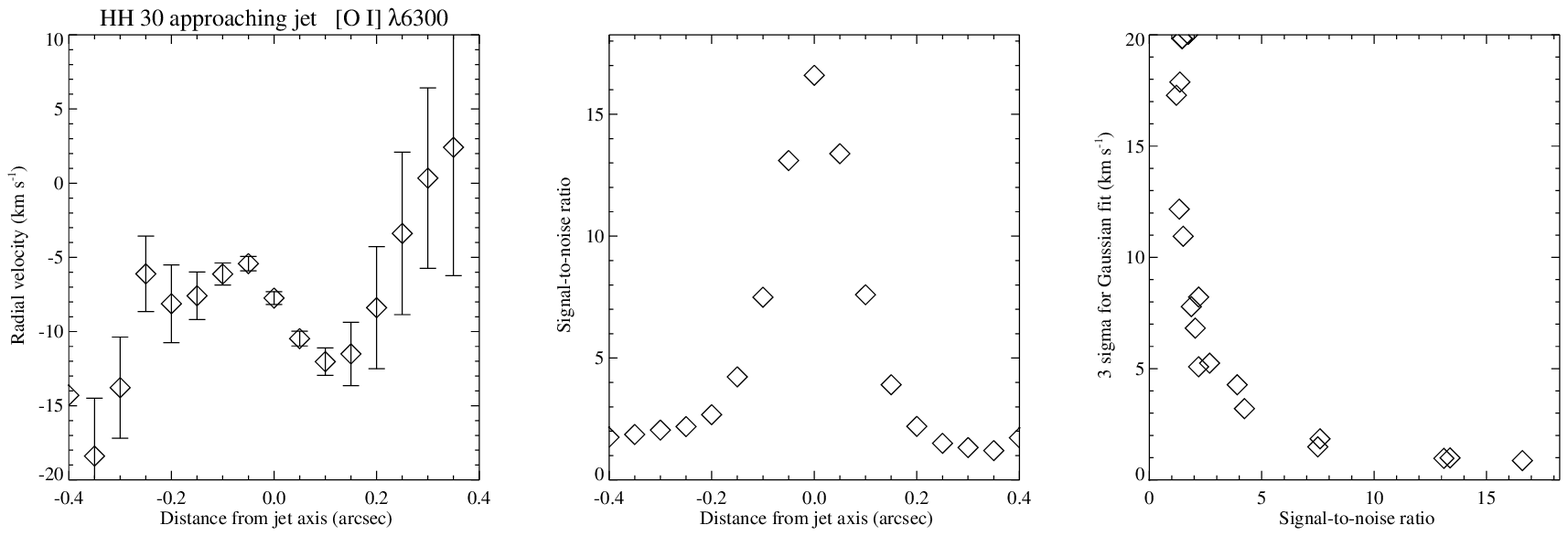}
\figcaption{Left - Velocity profiles for selected target and emission lines plotted with error bars representing the three-sigma error associated with each Gaussian fit; Centre - signal-to-noise of binned intensity profile after emission is centered on the jet axis; Right - variation of signal-to-noise with errors on Gaussian fits. Fitting becomes unreliable typically below a signal-to-noise ratio of 3. 
\label{errorbars_opt_nuv}}
\end{center}
\end{figure*}

Figure\,\ref{errorbars_opt_nuv} is shown as an illustration of the effect of a decrease in signal-to-noise on the three-sigma error bars associated with each Gaussian fit using IDL. Data points below signal-to-noise of 3 were generally deemed unreliable. Given 
sufficient signal-to-noise, we are able to routinely measure radial velocity differences to one fifth of the velocity 
sampling, conservatively speaking. This yields an accuracy of $\pm$5\,km\,s$^{-1}$ in the optical and $\pm$1\,km\,s$^{-1}$ in the NUV. However, since additional 
uncertainties are introduced by the NUV absorption feature, we prefer to give the conservative error estimate of 
$\pm$5\,km\,s$^{-1}$ for our NUV results. 

Meaningful analysis of much fainter NUV emission lines evident in the observations 
was not possible. One of these emission peaks was identified, in the DG Tau jet spectrum, as the combined emission lines of 
\ion{C}{2}]\,$\lambda$$\lambda$2324,2325,2326,2327,2328 travelling at 180\,km\,s$^{-1}$. The other emission peak appears to be [\ion{O}{2}]\,$\lambda$2470\,, again if a jet radial velocity of about 180\,km\,s$^{-1}$ is assumed. 

\begin{figure*}
\begin{center}
\includegraphics[scale=0.9]{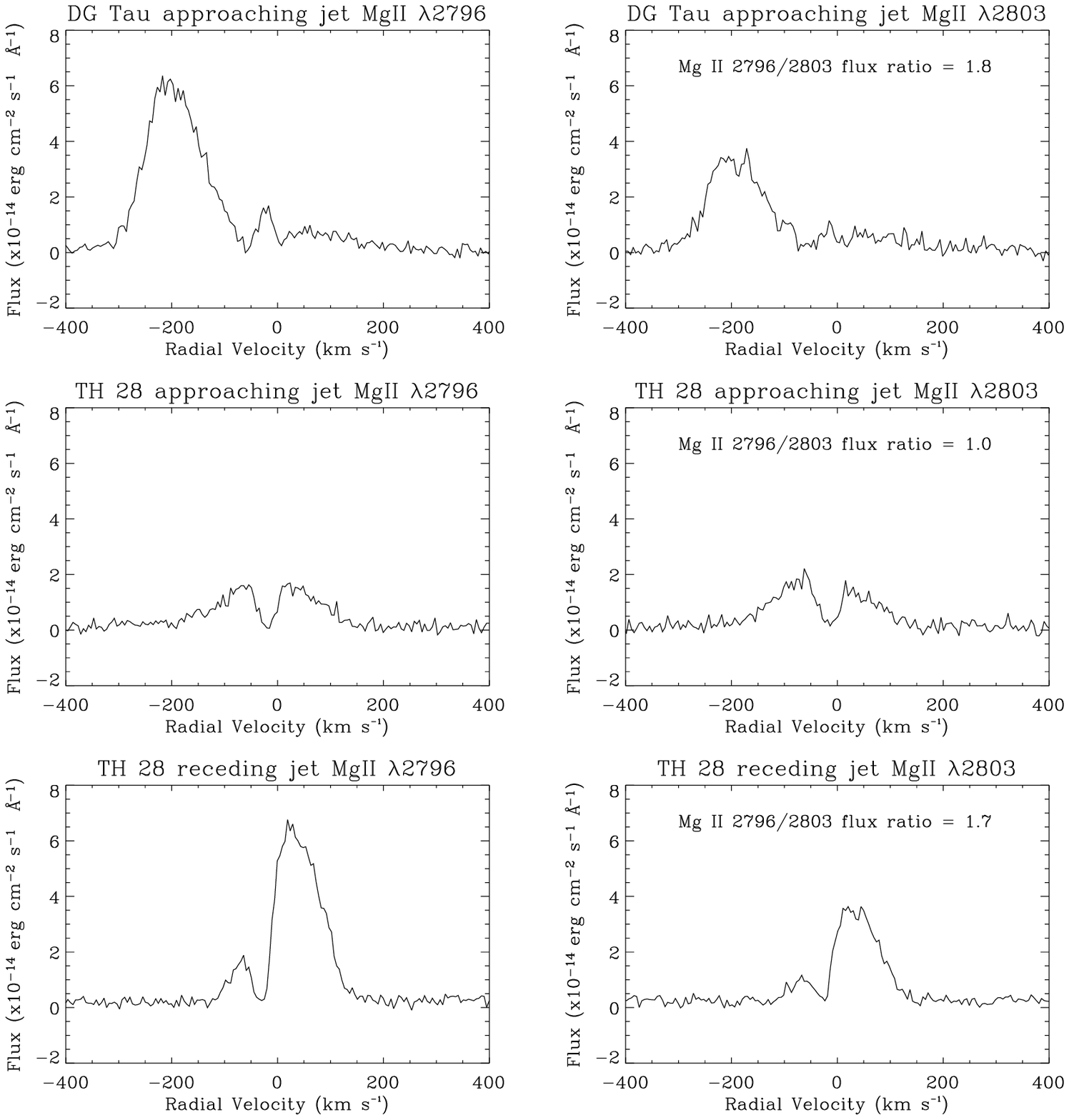}
\figcaption{Flux profiles of the \ion{Mg}{2}\,$\lambda$$\lambda$2796,2803 emission doublet, integrated across the jet, for 
the approaching jet from DG\,Tau and each lobe of the bipolar jet from TH\,28. The velocity scale is corrected for systemic 
heliocentric radial velocity, $v_{sys}$, Table\,\ref{targets}. Fluxes are quoted without compensation for the absorption feature, as the exact emission profile shape is not known. 
\label{intensityprofiles_uv}}
\end{center}
\end{figure*}

The \ion{Mg}{2} absorption feature dominates the jet emission from TH\,28, more so than in the case of DG\,Tau. This becomes evident when emission is binned for plotting in one dimension, Figure\,\ref{intensityprofiles_uv}. The ratio of \ion{Mg}{2} fluxes integrated across each jet for the entire emission line is given in each plot. Fluxes are quoted without extinction correction, which could not be determined from these spectra because pairs of transitions, from the same upper level, for any given species are of insufficient separation in wavelength to allow extinction calculation. Also, no compensation has been made for the absorption feature since the exact emission profile shape is not known. The absorption dip is located at low blue-shifted velocities for all three targets. Approximate speeds were measured as -26\,km\,s$^{-1}$ in the TH\,28 receding jet, -15\,km\,s$^{-1}$ in the TH\,28 approaching jet and -43\,km\,s$^{-1}$ in the DG\,Tau approaching jet (i.e. with repect to the vacuum wavelength but with no adjustment for the systemic velocity). 

\begin{table*}
\begin{center}
\scriptsize{\begin{tabular}{llccccccc}
\tableline\tableline

Target		&Distance		&$\Delta v_{rad}$ for	 	&$\Delta v_{rad}$ for 		&$\Delta v_{rad}$ 
for		&$\Delta v_{rad}$ for	&$\Delta v_{rad}$ for	&$\Delta v_{rad}$ for		&$\Delta v_{rad}$ for	\\
		&from axis		&[O I]$\lambda$6300		&[O I]$\lambda$6363 		&[N II]$\lambda$6548	
	&[N II]$\lambda$6583	&He I$\lambda$6678	&[S II]$\lambda$6716 	 	&[S II]$\lambda$6731	\\
		&(arcsec)		&(km~s$^{-1}$)			&(km~s$^{-1}$)			&(km~s$^{-1}$)		
	&(km~s$^{-1}$)		&(km~s$^{-1}$)		&(km~s$^{-1}$)			&(km~s$^{-1}$)		\\ 
\tableline
DG\,Tau		&0.05			&15 (14)			&19 (21)			&14				&7			&2			&5				&4				\\
approaching 	&0.10			&20 (17)			&22 (23)			&18				&5			&...			&4				&7				\\
jet		&0.15			&26 (18)			&13 (29)			&...				&7			&...			&-1				&9				\\
		&0.20			&10 (-2)			&...				&...				&...			&...			&...				&-4				\\
		&0.25			&17 (-3)			&...				&...				&...			&...			&...				&-3				\\
&&&&&&&&\\
CW\,Tau		&0.05			&15 (11)			&...				&...				
&12			&...			&...				&1 (3)				\\
approaching	&0.10			&15 (0)				&...				&...				
&15			&...			&...				&-5 (4)				\\
jet		&0.15			&12 (5)				&...				&...				
&...			&...			&...				&... (-1)				\\
&&&&&&&&\\

HH\,30		&0.05			&5				&...				&...				&0			&...			&-1 				&-1				\\
approaching	&0.10			&5				&...				&...				&3			&...			&-1 				&-2				\\
jet		&0.15			&4				&...				&...				&...			&...			&-3 				&-3				\\
		&0.20			&1				&...				&...				&...			&...			&...				&-4				\\
		&0.25			&...				&...				&...				&...			&...			&...				&-2				\\
\tableline
\end{tabular}}
\end{center}
\caption{Radial velocity differences, $\Delta v_{rad}$, between one side of the jet axis and the other measured 
using single Gaussian fitting. Only where the emission is divided into a clear higher and lower velocity component was double Gaussian fitting used. In these cases, the results for the lower velocity component are given in brackets beside 
the results for the higher velocity component. Where blanks appear in the table, the emission was either 
Doppler shifted off the CCD, or was too faint to decipher. The H$\alpha$ line from the jet was contaminated by reflected emission from the star thus rendering it too difficult to analyse. Finally, the [\ion{O}{1}]$\lambda$6363 line was often contaminated by the signal from a defective pixel. The error bars on the data analysis are at most $\pm$5 km~s$^{-1}$ but reduce with increasing signal-to-noise (see Figure\,\protect\ref{errorbars_opt_nuv}). A signal-to-noise cut-off ratio of 3 was used, as measurements below this level were generally deemed unreliable, which therefore excludes the other regions of the flow. 
\protect\label{rotational_velocities_optical}}
\end{table*} 
\begin{table*}
\begin{center}
\scriptsize{\begin{tabular}{llcc}
\tableline\tableline

Target			&Distance		&$\Delta v_{rad}$ for	&$\Delta v_{rad}$ for	\\
			&from jet axis		&Mg II\,$\lambda$2796	&Mg II\,$\lambda$2803	\\ 
			&(arcsec)		&(km\,s$^{-1}$)		&(km\,s$^{-1}$)		\\  
\tableline
DG\,Tau 		&0.029			&7			&2	 		\\
approaching		&0.058			&14			&6			\\
jet			&0.087			&14			&11			\\
			&0.116			&8			&...			\\ 
&&&\\
			
TH\,28 			&0.029			&0			&-3			\\
approaching		&0.058			&2			&...			\\
jet			&0.087			&1			&...			\\
&&&\\

TH\,28 			&0.029			&-3			&-1			\\
receding		&0.058			&-3			&-2			\\
jet			&0.087			&0	                &-4			\\
			&0.116			&1			&-3			\\
			&0.145			&1			&-2			\\
			&0.174			&3			&...			\\
			&0.203			&6			&...			\\ 
			&0.232			&3			&...			\\ 
\tableline
\tableline
\end{tabular}}
\end{center}
\caption{Radial velocity differences, $\Delta v_{rad}$, between one side of the jet axis 
and the other, measured in NUV lines using a cross-correlation technique. Cross-correlation was found to be more suitable than Gaussian fitting given that the profiles include disruptive absorption. Since the velocity profiles of Figure\,\protect\ref{velocityprofiles_optuv} rely on Gaussian fitting, they should not be expected to exactly match this table, but are intended only as an indication of the transverse velocity profile. Where blanks appear in the table, the emission was too faint to decipher. The accuracy reached with the data analysis was approximately $\pm$5\,km~s$^{-1}$. A signal-to-noise cut-off ratio of 3 was used, as measurements below this level were generally deemed unreliable (Figure\,\protect\ref{errorbars_opt_nuv}), which therefore excludes the other regions of the flow. 
\label{rotational_velocities_uv}}
\end{table*}
\section{Discussion}
\label{discussion}

\subsection{Indications of Jet Rotation}
\label{Indications_of_Jet_Rotation}

In almost all cases, the results show clear gradients in Doppler shift across the jet base, in the optical and NUV wavelength regions. One interpretion of these radial velocity differences is as indications of jet rotation close to the source where the jet is launched. 

Position-velocity diagrams for the optical forbidden emission lines give a qualitative indication of rotation in 
the form of a tilt in the contours, Figure\,\ref{pv_optical_jets}. Furthermore, from these diagrams and the NUV flux profile plots (Figure\,\ref{intensityprofiles_uv}) we see how different emission lines trace different velocities within the jet: [\ion{N}{2}] lines appear to trace only higher velocities; \ion{Mg}{2} and [\ion{O}{1}] lines trace both higher and lower velocities; and [\ion{S}{2}] lines trace mainly lower velocities. The diagrams also show how the jet becomes spatially resolved at low velocities, as is clearly evident in cases where the emission traces both higher and lower velocity material, i.e. the DG\,Tau [\ion{O}{1}] and [\ion{S}{2}] emission and the CW\,Tau [\ion{O}{1}] emission. Meanwhile, for all targets, the [\ion{S}{2}] lines show this spatial extension but little or no rotation, in the sense that radial velocity differences mainly lie within the error bars of 5\,km\,s$^{-1}$ about zero (Figure\,\ref{velocity_shifts}). This may be due to the fact that at a given distance from the jet axis, line-of-sight and projection effects (as discussed in (Pesenti\,et\,al.\,2004)) reduce the measured Doppler shift more considerably in the case of the broader [\ion{S}{2}] flow than, for example, the [\ion{O}{1}] flow. 

The transverse radial velocity profiles for all targets (Figures\,\ref{velocityprofiles_opt} and \ref{velocityprofiles_optuv}) further highlight the asymmetry in the radial velocities across the jet. The clearest cases are those of the DG\,Tau approaching jet and the TH\,28 receding jet. Here the plots reveal that the on-axis jet material is travelling fastest while the borders of the jet are travelling slower. The result is a v-shaped profile. But these profiles are not symmetrical: velocities on one side of the axis are lower than on the other, a natural outcome if the jet is rotating. In the case of the CW\,Tau higher velocity jet profile, the emission is more confined. Its profile does not incorporate a significant border contribution, and it is clearly separated from lower velocity material (Figure\,\ref{pv_optical_jets}). It therefore does not trace an asymmetric v-shape, but instead exhibits a simple slope whereby velocities on one side of 0$\arcsec$ are lower than on the other side. It is natural in such cases that the central datapoint should not be the highest radial velocity point, but should have an intermediate velocity value (within error bars). Examining all target profile plots, and allowing for an uncertaintly of 5\,km\,s$^{-1}$, the variety within the plots can be explained according to one or both of these two basic profile shapes. The combined profiles of the 
optical and NUV datasets illustrate how measured radial velocities are comparable for \ion{Mg}{2} and optical lines for both the 
approaching jet from DG\,Tau and the receding jet from TH\,28. The approaching jet from TH\,28 does not follow this 
trend. The fact that this jet is inclined at only 10\degr\,\,with respect to the plane of the sky (Table\,\ref{targets}) 
means that it has low radial velocity values, and so the peak jet velocity almost coincides with the velocity of the absorption 
feature. This prevents reliable estimates of radial velocity from NUV lines for this target. 

Measured radial velocity differences, Tables\,\ref{rotational_velocities_optical} and \ref{rotational_velocities_uv}, show values typically in the range of 10~to~25\,km\,s$^{-1}$ for DG\,Tau and CW\,Tau in the optical region within 0.$\arcsec$25 of the jet axis. Meanwhile, marginal differences were detected for HH\,30 of typically 5\,km\,s$^{-1}$, which lie close to or within the error bars about zero. In the NUV region, the \ion{Mg}{2} doublet shows radial velocity differences for DG\,Tau of about 5\,to\,15\,km\,s$^{-1}$ within 0.$\arcsec$1 of the jet axis which is in line with, but notably lower than, the optical values. The values for TH\,28 are all within error bars about zero, except for a measurement of 6\,km\,s$^{-1}$ at 0.$\arcsec$2 from the jet axis in the receding jet. Note that in a previously published paper, optical values of 10~to~24\,km\,s$^{-1}$ were measured for the same region \protect\cite{Coffey04}. Difficulties in tackling the absorption dip possibly contribute to lack of detectable velocity differences in this NUV dataset. 

To comment generally on Tables\,\ref{rotational_velocities_optical} and \ref{rotational_velocities_uv}, there appears to be a drop in velocity differences close to the jet axis for DG\,Tau in the NUV. Such lower values close to the axis were also noted for other jets in the optical analysis of Coffey~et~al.~(2004) and outlined 
as being most likely due to line-of-sight and projection effects (\protect\citealp{Pesenti04}; \protect\citealp{Dougados04}). 
However, this drop-off does not appear in the optical results for DG\,Tau to the same extent. Also, Coffey~et~al.~(2004) describes [\ion{N}{2}] lines as showing higher rotation since this emission line traces higher 
velocities, but this is not supported here in the optical results of DG\,Tau and CW\,Tau, or the NUV results where 
the \ion{Mg}{2} doublet traces equally high velocities as [\ion{N}{2}] emission. 

Interpreting these gradients in Doppler shift as indications of jet rotation reveals a sense of rotation as indicated in Figure\,\ref{orientation}. Supporting this interpretation is the fact that there is agreement in direction of Doppler gradient between the DG\,Tau optical and NUV results. For the TH\,28 receding jet, there is no clear gradient detected in the NUV, although the outlier of 6\,km\,s$^{-1}$ at 0\farcs2 would suggest the same sense of rotation as measured in previously published optical data for this target \protect\cite{Coffey04}. 

A large velocity difference was not measured for HH\,30. This is surprising, given that this system is in the plane of the sky and so should present the most easily measureable transverse Doppler gradient. Interpreting the differences as rotation would imply that the approaching jet is rotating clockwise, looking down the approaching jet towards the star. This finding would conflict with the measured sense of disk rotation \protect\cite{Pety06}. However, since these measurements are marginal, it appears that we cannot conclusively derive a sense of rotation for the jet. Meanwhile, no rotation signature was detectable in the surrounding CO outflow associated with this optical jet \protect\cite{Pety06}. 

An alternative explanation for a gradient in radial velocity across the jet base could be the presence of asymmetric shocks, since shocks are invoked to explain jet line emission. Indeed it is conceivable that we are observing a combination of effects, including both jet rotation and asymmetric shocking of jet material. However, observations of Doppler gradients across the bipolar jet of RW\,Aur taken with several slits parallel to the jet axis \protect\cite{Woitas05} demonstrate that if an asymmetric shock was causing these gradients, the shock velocity asymmetry would have to persist along the jet for at least $\sim$\,200\,AU to explain these observations. Not only that, but the same shock asymmetry would have to persist {\em over this distance} in {\em both lobes} simultaneously. Further observations of Doppler gradients in bipolar systems (\protect\citealp{Coffey04}), which also reveal agreement in the direction of the gradient within the bipolar jet, demonsatrate that the direction and magnitude of the shock velocity asymmetry would have to be the same in each lobe in the bipolar system. Therefore, while asymmetric shocking is a conceivable explanation, it becomes less likely within the context of these results. 

Finally, Cerqueira et al.\ (2006) have shown how the small-angle precession of a jet
can produce the transverse velocity differences seen here and hence mimic
rotation. In particular, they have modelled a jet using a mean velocity of 300\,km\,s$^{-1}$ along with a precessional angle and period of 5$^{\circ}$ and 8\,years respectively. Some of the jet parameters are based on the DG\,Tau jet but, for example, they choose a jet inclination angle of 45$^{\circ}$ rather than the observed value for DG\,Tau of 37.7$^{\circ}\pm$\,2 \protect\cite{Eisloffel98}. For their simulations they also assume that the jet has a time-dependent ejection velocity that varies with the
same period as the precession. In the case of DG\,Tau, there is ample evidence
from its morphology that the jet is precessing (Lavalley-Fouquet
et al.\ 2000; Dougados et al.\ 2000). Thus for DG\,Tau a
precessional model may be reasonable; although this does not, of course,
exclude the presence of rotation as well. That such a model is applicable
however {\em in every case} in which we see transverse velocity
differences of a few tens of km\,s$^{-1}$ seems contrived. For example, if
the precessional angle were only 1-2$^{\circ}$, instead of 5$^{\circ}$,
no measurable effect would be seen in their pure precessional model of the
DG\,Tau jet. Likewise the precessional period, etc., are critical.
Moreover, as Cerqueira et al.\ (2006) remark themselves, there is no evidence
in some cases, e.g. the RW\,Aur jet, for precession. That said,
any ambiguity as to the cause of the transverse velocity differences in the
case of individual outflows can, and should, be removed through modelling. In
particular precessional parameters can be obtained from multi-epoch imaging
and fitting of jet emission centroids as a function of distance from the
source. 

\subsection{Error Analysis}
\label{error_analysis}

With reference to errors, it should be mentioned that, during the first stage of data reduction (see the beginning of Section\,\ref{Quantitative_analysis}), when each emission peak was recentred to the middle of the pixel in the central row of the detector (i.e. the nominal 0\arcsec position on the detector), a trend was observed in the pixel offsets from the 0\arcsec position 
for each emission line. Peak offsets ranged from 0.1 to 1.5 pixels depending on the target, although offsets in 
emission lines for a given target varied by $<$0.5 pixels. This highlighted three types of possible instrument 
misalignments, which had the potential to contribute to position-velocity contour tilt and thus mimic rotation. The 
first is physical tilt (i.e. a tilt of the slit with respect to the nominal observing position angle), and has 
values of +0.22 degrees in the optical region and +0.27 degrees in the NUV. The second is optical distortion tilt 
(i.e. the slit image on the detector is curved and tilted) which varies in angle across the detector depending on 
the grating used, and has values of +0.001 degrees in the optical region and +0.9 to +1.2 degrees in the NUV. The third is position angle error (i.e. a difference between the slit position angle used in the observations with respect to that requested), and could be on the order of degrees. For example, as noted in Figure\,\ref{orientation} the actual slit position angles for DG\,Tau and HH\,30 differ from the requested values by +3$^{\circ}$ and -6$^{\circ}$ respectively. This was due to problems during observations in finding the right combination of guide stars for the requested slit position angle. 

All possible corrections for errors arising from instrument misalignment were made before data analysis was conducted. For the optical data, the {\em HST} pipeline calibration accounted for physical tilt. The optical distortion, on the other hand, was not removed in the pipeline. Moving from right to left along the CCD, the spectrum gradually curves upwards with a total difference of half a pixel from one end to the other. In other words, the [\ion{O}{1}] lines are raised higher on the detector than the [\ion{S}{2}] lines. This was adequately accounted for by recentering the emission peaks in the middle of a pixel. There could still remains elements of the distortion which cannot be accounted for, since its exact nature is unknown, but the effect on measurements cannot be significant. A confirmation of this is given by {\it HST}/STIS archival data with almost the same instrument configuration as our observations. A long-slit spectrum of Mz\,3 was obtained with {\it HST}/STIS on June 23, 2002 (Proposal ID 9050). The same grating (G750M) and detector (CCD) were used, but the slit aperture size used was slightly different (i.e. 52$\times$0.05\,arcsec$^{2}$ rather than 52$\times$0.1\,arcsec$^{2}$). These slits are in fact coaligned and their optical paths are the same, so this difference in aperture size does not affect our investigation. The archival spectrum shows flux levels comparable in intensity and spatial extension to our data, and exhibits [\ion{S}{3}] emission which hits the detector in the region between the two components of the [\ion{O}{1}] doublet. Reassuringly, no tilt is present in the [\ion{S}{3}] line. Furthermore, Gaussian fitting to pixel rows either side of the 0\arcsec position for the [\ion{S}{3}] line revealed velocity differences of effectively zero (ie. $\leq$0.5\,km\,s$^{-1}$). Given the extreme unlikelihood that instrumental effects could operate in this manner (i.e. causing a tilt in the [\ion{O}{1}] doublet but no tilt in emission between the doublet lines), this data confirms the fact that the instrument is not introducing spurious contour tilts in our data. 

For the NUV data, standard IRAF routines were used to wavelength calibrate the data. These routines rectify the spectrum both spatial and spectrally, although without accounting for the optical distortion. Therefore, prior to this rectification, the calibration lamp lines were used to determine the extent of the optical distortion tilt in the region on the detector of interest. The STIS calibration reference tables were then manually adjusted to incorporate the value of this tilt angle prior to wavelength calibration. In this way, standard routines accounted for both physical tilt and optical distortion. As in the case of the optical data, any residual effect cannot be accounted for since the nature of the distortion is unknown. Nevertheless, it cannot significantly affect our measurements since, for DG\,Tau, our NUV measurements are in reasonable agreement with the optical measurements even though they were taken with different instruments and have a different optical distortion to contend with. 

The remaining tilt to be addressed is possible inaccuracy in the position angle specified for observations. Although this is mainly corrected for through recentering the emission peaks to the centre of a pixel (as described in the beginning of Section\,\ref{Quantitative_analysis}), it is possible that a tilt error could remain unaccounted for. The precise effect of this misalignment is not known, since the kinematic nature of the underlying flow is not known. However, in the simplest case of a cylindrical, non-accelerating, non-rotating flow a slit misalignment does not introduce a radial velocity gradient across the jet. Therefore, for our observations, we consider the effect insignificant. 

The fact should be noted that we see the {\em same sense} of Doppler gradient {\em with respect to the slit orientation} for all T\,Tauri jets in the optical 
and NUV survey for which a Doppler gradient is detected. This includes previously published optical results for both lobes of the bipolar jets from TH\,28 and 
RW\,Aur \protect\cite{Coffey04}. (Note that the previously published results for LkH$\alpha$\,321 have since proved inconclusive.) This fact might be regarded as unusual, and could point to instrumental error. Analysis of archival Mz3 [\ion{S}{3}] emission (as described above) excludes this possibility. Furthermore, bare in mind that, of the eight jet lobes, the seven that show Doppler gradients (i.e. excluding LkH$\alpha$\,321) are each launched from one of five 
T\,Tauri systems. This effectively reduces the sample size, making the statistic more acceptable. It should be noted that a  
rotation analysis of the radial velocities within the jets from T\,Tauri stars DG\,Tau (Bacciotti~et~al.~2002) 
and RW\,Aur (Woitas~et~al.~2004) conducted in the optical region with the slit placed {\em parallel} to the jet axis yielded results in 
agreement with our observing mode in which the slit is {\em perpendicular} to the jet axis. This strongly negates instrumental error. To further strengthen the case, 
it is important to also highlight the fact that the observations for the jets from both DG\,Tau and TH\,28 in 
the optical and NUV regions were conducted with {\em different detectors} (and so different optics). For DG\,Tau the results are in good agreement, while for TH\,28 the results are not inconsistent. This combination of studies, along with the careful analysis of 
instrumental effects, strongly suggests that, although it seems fortuitous that all jets (with the possible 
exception of LkH$\alpha$\,321) have Doppler gradients in the same direction with respect to the slit orientation, it is not 
impossible and does indeed seem to be the case. Finally, note that four of the five T\,Tauri systems 
with Doppler gradients are located in the same cloud (i.e. Taurus-Auriga), although no particular pattern of 
system alignment could be readily identified. 

\subsection{\ion{Mg}{2} Absorption}
\label{MgII_absorption}

\begin{figure}
\begin{center}
\includegraphics[scale=0.44]{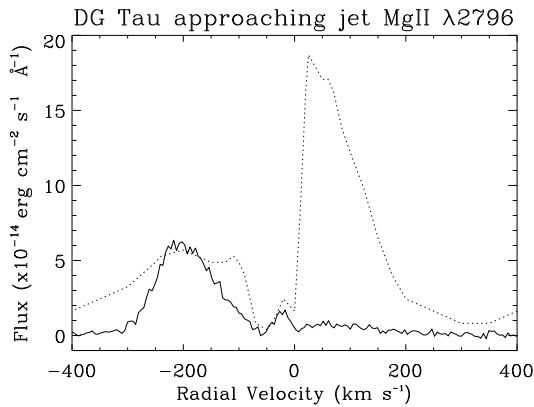}
\figcaption{\ion{Mg}{2}\,$\lambda$2796 flux profile for the DG Tau system. The solid line represents our observation of the approaching jet at 0$\farcs$3 from the star, taken from Figure\,\ref{intensityprofiles_uv}. The dashed line represents observations in 1996 of a 1.74 arcsec squared area centered on the star, extracted from Figure 5 of \protect\citealp{Ardila02}. The comparison serves to illustrate that the outflow is both absorbing and {\em emitting} in \ion{Mg}{2}, and also that the stellar PSF does not enter the slit in our observations. 
\label{ardila}}
\end{center}
\end{figure}

In recent studies, several spectra of T\,Tauri stars have been obtained, in which the associated jet was {\em included} in the slit coverage (\protect\citealp{GomezdeCastro01}; 
\protect\citealp{Ardila02}). Ardila\,et\,al.\,(2002), in particular, present a spectrum of DG\,Tau taken with the Goddard High Resolution Spectrograph onboard {\em HST}. The coverage was $\sim$2$\arcsec$ centered on the star, and so the spectrum 
included both the magnetospheric accretion region and the base of the jet. The authors assume that {\em all} the \ion{Mg}{2} emission 
originates from the magnetosphere, and attribute the broad blue-shifted dip to absorption by a cool outflow. Figure\,\ref{ardila} shows a comparison of their data and ours, and clearly demonstrates that the jet is both absorbing and {\em emitting}. Their spectrum includes emission of \ion{Mg}{2} radiation from the jet such that our spectrum corresponds both in velocity and intensity to the blue-shifted plateau reported in Ardila\,et\,al.\,(2002). In fact, one-dimensional NUV spectra similar to ours were obtained for the bow-shock HH\,47A with the Faint Object Camera on-board {\em HST} \protect\cite{Hartigan99}. In that case, the line intensities have been successfully modelled as arising from composite shocks advancing into a low density, and moderately ionised, medium. 
Figure\,\ref{ardila} also serves to illustrate that the stellar PSF does not enter the slit in our observations. Recall from Section\,\ref{results} that, for our observations of the jet base, the wings of the {\em HST} spatial line spread function in both the NUV and the optical regions do not extend as far as the star, and so there can be no contamination from photospheric line emission. This is supported in Figure\,\ref{ardila} by the lack of emission at intermediate velocities in our observations compared to observations centered on the star. The higher jet velocities ($>$ 250\,km\,s$^{-1}$) of Ardila\,et\,al.\,(2002) are not seen in our spectrum, which seems to indicate that our observations were taken at a distance from the star which is before the end of the jet acceleration zone. This is consistent with evidence of increasing velocity with distance along the DG\,Tau jet as reported by Bacciotti\,et\,al.\,(2000). Similarily, intermediate velocity emission ($\sim$\,100\,km\,s$^{-1}$) either originates from the wings of the stellar emission profile, or from jet material closer to the star which is yet to be fully accelerated. 

In our spectra, a low velocity absorption feature is also present within the emission profile of all three targets (Figure\,\ref{intensityprofiles_uv}). This dip is travelling at low blue-shifted velocities, has a narrower FWHM ($\sim$\,40\,km\,s$^{-1}$), and is not significantly asymmetric as would be expected from wind absorption. Since it is {\em blue}-shifted in all the observed targets, including the {\em receding} jet from TH\,28, it cannot originate in the jet lobe itself. The dip does coincides with interstellar absorption velocities, and so it is conceivable that it originates from interstellar cloud material. \ion{Mg}{2} interstellar absorption features are present, for example, in spectra of main~sequence F stars (B\"{o}hm-Vitense~et~al.~2001). If this were the case in our spectra, explanation of the absorption FWHM would necessitate the combined contribution of two or three interstellar clouds (see also discussion by Ardila\,et\,al.\,(2002)). This assumption of multiple interstellar clouds would have to be made in the direction of both DG\,Tau and TH\,28, making this interpretation less plausable. The feature could be more readily explained by an expanding shell around the T\,Tauri system. In contrast to attributing it to a cooler outflow wind, an expanding shell would be observed as a blue-shifted dip regardless of whether the observed jet is approaching or receding. A shell interpretation would also account for a FWHM which is possibly too large to be caused by ISM absorption. 

A final point concerns the flux intensity ratio \ion{Mg}{2}$\lambda$2796/$\lambda$2803 within the jet. Statistically, this ratio is expected to be 2:1 when emitted by an optically thin gas, while it approaches 1:1 in conditions of optical thickness (see e.g. \protect\citealp{bohm01}). Hence, the flux ratios (Figure\,\ref{intensityprofiles_uv}) for the TH\,28 receding jet and the DG\,Tau approaching jet demonstrate the gas to be optically thin. Meanwhile, the TH\,28 approaching jet is evidently optically thick, and the FWHM of the dip is slightly larger here than for the other targets. Therefore, in this case, the absorption feature is very likely to incorporate a self-absorption component. Besides, this jet is known to be heavily embedded in reflection nebulosity (Graham \& Heyer 1998). This combined contribution to the dip, along with the fact that the emission and absorption peaks overlap considerably, leads to particular difficulties in determining accurate radial velocities for emission from jet target. 

\subsection{Jet Launch Point}

Under the assumption that the radial velocity gradient measurements described in Section\,\ref{results} are to be interpreted soley as jet rotation, and that the mechanism for launching this portion of the jet can be accurately described by a magnetocentrifugal stationary wind, then the quantities derived in this work can be used to constrain the properties of the launch model. In particular, we can use our radial velocity measurements, in combination with the values in Table\,\ref{targets}, to 
find a range for the radius from the star on the disk plane of the jet footpoint (or launch point), $\varpi_{0}$. This was calculated using Equation\,\ref{foot_point} in 
Anderson et\,al.\,(2003), which is derived from general conservation properties of magnetocentrifugal models for jet acceleration, 
\begin{eqnarray}
\varpi_{0} & \approx & 0.7AU 
\left(\frac{\varpi_{\infty}}{10AU}\right)^{2/3}\ 
\left(\frac{v_{\phi,\infty}}{10kms^{-1}}\right)^{2/3}\ \nonumber \\ & & \times 
\left(\frac{v_{p,\infty}}{100 kms^{-1}}\right)^{-4/3}\ \left(\frac{M_{\star}}{1M_{\odot}}\right)^{1/3} 
\label{foot_point}
\end{eqnarray}

where $\varpi_{\infty}$ is the radius from the jet axis of the observation (conducted at an effectively infinite 
distance from the star), $v_{\phi,\infty}$ is the toroidal velocity at the observation radius, $v_{p,\infty}$ is 
the poloidal velocity at the observation radius, and $M_{\star}$ is the mass of the star. Since the inclination angle, $i_{jet}$, given in Table\,\ref{targets} is with respect to the plane of the sky, then $v_{\phi}=(\triangle v_{rad}/2)/\cos\,i$ and $v_{p}=(\overline{v_{rad}})/\sin\,i$, where $\overline{v_{rad}}$ is the average of the radial velocity measured at a given distance each side of the jet axis. The equation is valid for values of $v_{\phi} \gg v_{p}$, which is achieved at distances from the star of typically a few AU, and hence well before the position of our slit at several tens of AU. Where M$_{\star}$ is unknown, it is assumed to be equal to 1\,M${_\odot}$. 

The range for the jet footpoint radius for each target is given in Table\,\ref{footpoint}. Values were calculated for a sample of emission lines, rather than giving an average for each target, since each line traces different levels of radial velocity. Calculations were carried out for radial velocity differences measured far from the jet axis, to minimise line-of-sight and projection effects. These effects only allow measurement of lower limits in radial velocity within typically one spatial FWHM from the axis, a critical point for the determination of true toroidal velocities and hence footpoints (Pesenti\,et\,al.\,2004). 

From the results of the three emission lines for DG\,Tau, for example, we see that the higher velocity component appears to be launched from a distance of 0.2~to~0.5\,AU from the star along the disk plane, while the lower velocity [\ion{O}{1}] component appears to trace a wider part of the jet launched from as far as 1.9\,AU. These values are in the same range as those estimated for the DG\,Tau jet of 1.8\,AU by Bacciotti\,et\,al.\,(2002), using velocity differences from data taken with the STIS slit parallel to the jet axis, and of 0.3~to~4\,AU by Anderson\,et\,al.\,(2003) for the same dataset (where the analysis was carried out on only the lower velocity component of the jet). The results for the other targets also fall into this range, determining the footpoint of the jet to be within 0.2~to~3.9\,AU from the star. Note, also, that DG\,Tau presents a good case for examining the variation of poloidal velocity with footpoint radius. Here the low and high velocity jet material have comparable radial velocity differences, and so we can see that decreasing the poloidal velocities gives rise to an equivalent increase in footpoint radius. 

Given the uncertainties associated with many of the physical parameters (such as stellar mass, inclination angle and stellar distance), these calculations represent {\em estimates} of the jet footpoint and so no error bars are given. Nevertheless, the results support the idea that jets are launched, via the magneto-centrifugal mechanism, at footpoint radii within a few AU of the star. The results consistently show that, given our spatial resolution, the portion of the jet that we observe appears to be launched from a region up to a few AU from the star on the disk plane. However, this one-to-one mapping of the footpoint is probably an over simplification. In fact, MHD instabilities are likely to influence the geometry of the field lines beyond the Alfv\'{e}n surface, and so the jet material cannot necessarily be so easily traced back to exact footpoints on the disk plane. 

\begin{table*}
\begin{center}
\scriptsize{
\begin{tabular}{lcccccccc}
\tableline \tableline
Target				&Emission	&$\varpi_{\infty}$	&$\varpi_{\infty}$	&$\triangle v_{rad}$ 	
&$\overline{v_{rad}}$&$v_{\phi,\infty}$	&$v_{p,\infty}$	&$\varpi_{0}$ 	\\
				&line		&(arcsec)		&(AU)			&(km~s$^{-1}$)		
&(km~s$^{-1}$)	&(km~s$^{-1}$)		&(km~s$^{-1}$)	&(AU)		\\ 
\tableline
DG\,Tau approaching jet		&$\lambda$6300	&0.15			&21			&26 (18)		&195 (60)	&21 (15)		&248 (76)  	&0.5 (1.9)	\\
				&$\lambda$6583	&0.15			&21			&7			&165		&6			&209		&0.3		\\
		 		&$\lambda$2796	&0.116			&16			&8			&193		&6			&244		&0.2		\\ 
&&&&&&&&\\
CW\,Tau approaching jet		&$\lambda$6300	&0.15			&21			&12 			&108 		&8 			&164		&0.6		\\		
				&$\lambda$6583	&0.10			&14			&15			&102		&10			&155		&0.5		\\		
&&&&&&&&\\
TH\,28 receding jet		&$\lambda$6300	&0.20			&34			&16			&8		&8			&46		&3.9		\\
				&$\lambda$6583	&0.20			&34			&24			&27		&12			&155		&1.0		\\ 
				&$\lambda$2796	&0.20			&34			&6			&13		&3			&75		&1.0		\\ 
\tableline
\end{tabular}
}
\end{center}
\caption{
The radius from the star on the disk plane of the jet footpoint (or launch point), $\varpi_{0}$, calculated for targets in the 
optical and NUV using the method described in Anderson~et~al.~(2003). Values in brackets relate to the 
lower velocity component. The mean radial velocity, taken from values equidistant either side of the jet axis, 
$\overline{v_{rad}}$, is quoted as an absolute value (and after systemic heliocentric radial velocity correction). Toroidal and poloidal velocities were calculated as $v_{\phi}=(\triangle v_{rad}/2)/\cos\,i$ and $v_{p}=(\overline{v_{rad}})/\sin\,i$. The TH\,28 [\ion{O}{1}] and [\ion{N}{2}] results are taken from data reported in \protect\protect\cite{Coffey04}. 
\label{footpoint}}
\end{table*}

\section{Conclusions}
\label{conclusions}

In the context of a {\em HST}/STIS survey to ascertain the commonality of T\,Tauri jet rotation, 
we report on observations of three jet targets at optical wavelengths (i.e. the approaching jet 
from DG\,Tau, CW\,Tau and HH\,30), and three jet targets at NUV wavelengths 
(i.e. the bipolar jet from TH\,28 and the approaching jet from DG\,Tau). 
Almost all targets show distinct and systematic radial velocity asymmetries 
for opposing positions with respect to the jet axis, within 100\,AU from the source. 
Radial velocity differences of typically 10\,to\,25\,($\pm$5)\,km\,s$^{-1}$ were found at optical wavelengths, while values at NUV wavelengths were lower at typically 10\,($\pm$5)\,km\,s$^{-1}$. 
We interpret these radial velocity asymmetries as rotation signatures 
in the region close to the star where the jet has been collimated but has not yet 
manifestly interacted with the environment. 
Possible instrument contribution to error has been thoroughly 
examined and ruled out. We also note the presence of a \ion{Mg}{2} blue-shifted absorption feature in all three NUV cases. This possibly originates from the combined contribution of several interstellar clouds, or from an expanding shell of material around the T\,Tauri system. 

It was hoped that the higher spatial and spectral resolution of the {\em HST}/STIS in the NUV would afford a more quantitative analysis, in terms of an error bar reduction from one fifth of the spectral sampling in the optical region (i.e. $\pm$\,5\,km\,s$^{-1}$) to that of the NUV (i.e. $\pm$\,1\,km\,s$^{-1}$). However, we did not expect to find the tricky combination of a broad profile shape in the dispersion direction, a relatively narrow spatial FWHM, and significant interruption by absorption. This forced us to adopt a conservative approach to errors, i.e $\pm$\,5\,km\,s$^{-1}$. Nevertheless, in the case of DG\,Tau, the NUV measurements fit well with the results obtained from the optical data, and there is agreement in magnitude and direction of the Doppler gradients within the two wavelength ranges. The results for TH\,28 were, unfortunately, less clear and no sense of Doppler gradient could be established with certainty. Finally, the survey results are consistent with previously published results for DG\,Tau \protect\cite{Bacciotti02}. 

The determined rotational velocities were shown to be in agreement with magneto-centrifugal launch model predictions (e.g. Anderson~et~al.~2003). In the case of DG\,Tau, for example, we see that the higher velocity component appears to be launched from a distance of 0.2~to~0.5\,AU from the star along the disk plane. Meanwhile the lower velocity component appears to trace a wider part of the jet launched from as far as 1.9\,AU. The results for the other targets are similar and consistently show that, at the resolution of the observations, the jet is launched from a region of up to 3.9\,AU from the star on the disk plane. Therefore, if indeed the detected Doppler gradients trace rotation within the jet then, under the assumption of steady MHD ejection, the derived footpoint radii support the existence of magnetized disk winds. However, since we do not resolved the innermost layers of the flow, we cannot exclude the possibility that there also exists an X-wind \protect\cite{Shang07} or stellar wind component \protect\cite{Pudritz07}. Although we cannot probe the inner axial part of the flow at current resolution, this may be possible in the near future via interferometry. 

\vspace {0.2in}
{\bf Acknowledgements} 
\vspace {0.1in}
\newline
D.C. and T.P.R. would like to acknowledge support for their research from 
Enterprise Ireland and Science Foundation Ireland under contract 04/BR6/PO2741. J. E. and J. W. 
likewise wish to acknowledge support from the Deutsches Zentrum 
f\"ur Luft- und Raumfahrt under grant number 50 OR 0009. 
The present work was supported in part by the European Community's Marie Curie Actions - Human Resource and Mobility within the JETSET (Jet Simulations, Experiments and Theory) network, under contract MRTN-CT-2004-005592. 


\end{document}